\documentclass[%
 reprint,
 amsmath,amssymb,longbibliography,
 aps,
 prl,
]{revtex4-2}
\usepackage{graphicx}
\usepackage{dcolumn}
\usepackage{bm}
 \usepackage{color}
 \usepackage{tabularx}
\usepackage{tikz}
\usepackage[labelfont=bf]{caption} 
\usepackage[compat=1.1.0]{tikz-feynman}

\usepackage{ragged2e}

\usepackage{hyperref}

\usepackage{graphicx}
\usepackage{amsmath}
\usepackage{rotating}
\usepackage{color}

\newcommand{\sruo}{Sr$_2$RuO$_4$}
\newcommand{\lsoc}{\lambda_{soc}}
\newcommand{\io}{\tilde \mu}
\newcommand{\inu}{\tilde \nu}

\newcommand{\spin}{\sigma}

\newcommand{\pv}{{\bf p}}
\newcommand{\kv}{{\bf k}}
\newcommand{\qv}{{\bf q}}
\newcommand{\Qv}{{\bf Q}}

\usetikzlibrary{arrows,positioning} 
\tikzset{
    photon/.style={decorate, decoration={snake}, draw=black},
    electron/.style={draw=black, postaction={decorate},
        decoration={markings,mark=at position .55 with {\arrow[draw=black,thick]{>}}}},
    gluon/.style={decorate, decoration={snake},draw=black}, 
    >=stealth',
    punkt/.style={
           rectangle,
           rounded corners,
           draw=black, very thick,
           text width=6.5em,
           minimum height=2em,
           text centered},
    pil/.style={
           ->,
           thick,
           shorten <=2pt,
           shorten >=2pt,}
}

\begin{document}

\title{Superconducting state of \sruo ~in the presence of longer-range Coulomb interactions}

\author{Astrid T. R\o mer$^1$, P. J. Hirschfeld$^2$, and Brian M. Andersen$^1$}
\affiliation{%
$^1$Niels Bohr Institute, University of Copenhagen, Jagtvej 128, DK-2200 Copenhagen, Denmark\\
$^2$Department of Physics, University of Florida, Gainesville, Florida 32611, USA}

\date{\today}

\begin{abstract}

The symmetry of the superconducting condensate in \sruo ~remains controversial after nuclear magnetic resonance (NMR) experiments recently overturned the dominant chiral $p$-wave paradigm.   
Several theoretical proposals have been put forward to account for existing experiments, including a $d+ig$-wave admixture, conjectured to be stabilized by longer-range Coulomb interactions.  
We perform a material-specific microscopic theoretical study of 
pairing by spin- and charge-fluctuations in \sruo, including the effects of spin-orbit coupling, and both local and longer-range Coulomb repulsion. The latter has important consequences for \sruo~due to the near-degeneracy of symmetry-distinct pairing states in this material. We find that both the $g$- and $d_{x^2-y^2}$-wave channels remain noncompetitive compared to leading nodal $s'$, $d_{xy}$, and helical ($p$) solutions. This suggests nodal time-reversal symmetry broken $s'+id_{xy}$ or $s'+ip$ phases, promoted by longer-range Coulomb repulsion, as the most favorable candidates for \sruo.
We analyse the properties of these states, and show that the $s'+id_{xy}$ solution agrees with the bulk of available experimental data, including recent discoveries from NMR, muon spin relaxation ($\mu$SR), and ultrasound measurements.
\end{abstract}
\maketitle

{\it Introduction.} 
\sruo ~is in many respects an unconventional superconductor simple enough that  we should be able to understand why and how electrons pair. Superconductivity condenses around 1K from a well-characterized Fermi liquid, and samples are very clean.
The electronic structure is nearly two-dimensional (2D), with three bands at the Fermi level that mix $d$-orbital content only weakly, and spin-orbit coupling (SOC) is relatively small. Yet after two decades of study, its pairing state is unknown, in part because  a compelling narrative of a chiral $p$-wave ground state, based on early experiments, \cite{Mackenziereview,Sigrist2005,Maenoreview,Kallin2012,Mackenzie2017} dominated the scientific discussion.  This picture now seems unlikely after recent NMR experiments argued not only that the in-plane Knight shift is strongly reduced for temperatures below $T_c$\cite{Pustogow19,Ishida_correct}, but that the upper bound on the condensate spin-triplet magnetic response is at most 10\% of the normal state value\cite{Chronister}. 

If these experiments are borne out, the ground state of \sruo ~is a dominant spin-singlet pair condensate (some admixture of spin-triplet pairing must be present due to SOC). The difficulty with such a suggestion, however, is that the available pairing channels in tetragonal crystal symmetry of even parity are one-component representations with the exception of the $d_{xz}/d_{yz}~ (E_g)$ gap, which requires a pronounced three-dimensionality of the electronic structure\cite{suh2019}, unlikely in the case of \sruo. Nevertheless, a two-component solution is required to explain recent  resonant ultrasound spectroscopy~\cite{ghosh2020thermodynamic} and ultrasound velocity measurements~\cite{benhabib2020jump} as well as
measurements supporting time-reversal symmetry breaking (TRSB) in the ground state\cite{Luke1998,Kapitulnik09,grinenko2020split}. One way to obtain a two-component solution of dominant spin-singlet character is to combine two even-parity representations in a complex admixture, requiring an accidental degeneracy.  While this possibility was initially considered inelegant compared to scenarios with 2D representations, evidence in favor has been provided recently  by $\mu$SR experiments, which found that applied strain splits $T_c$ and the temperature at which TRSB  occurs\cite{grinenko2020split}.  

Of course, such states must satisfy constraints imposed by a wide variety of other experiments.  Specific heat, scanning tunneling microscopy (STM), and thermal transport experiments imply that low-energy quasiparticle states are present\cite{NishiZaki,Izawa_2001,Bonalde00,Firmo2013,Hassinger17,Kittaka2018,sharma20,Hassinger17}, a strong constraint since  accidental combinations of two singlet channels where the nodal structure of each is distinct will generically lead to a fully gapped state (e.g. $s+id$, if $s$ is nodeless).  

It is important to note, however, that nodal character deduced from low-order Fermi surface or Brillouin zone harmonics may be misleading. Realistic studies of unconventional superconductors, including Fe-based and heavy-fermion systems, often find ground states corresponding to higher-order harmonics because of strong momentum dependence of
the interaction driven by orbital physics, together with the need to minimize the local Coulomb energy (see e.g. Refs.~\onlinecite{Kreisel2013,Nomoto2016,Bjornson2021}). In the case of \sruo, several theoretical works have revealed a very rich superconducting gap structure\cite{Koikegami2003,Annett2006,Raghu2010,Wang2013,Scaffidi2014,Zhang18,Gingras18,Wang19}. In recent studies of spin-fluctuation-mediated superconductivity, we showed that \sruo ~exhibits a near-degeneracy of several different superconducting states, making the case of accidental degeneracy less unlikely\cite{RomerPRL,rmer2020fluctuationdriven,Romer_strain2020}.  In particular, states labelled $s'$ (A$_{1g}$) and $d_{x^2-y^2}$ (B$_{2g}$) were shown to be nearly degenerate over much of the phase diagram. An $s'+id_{x^2-y^2}$ state is nodal despite combining channels of $s$ and $d$ symmetry, because the $s'$ state has coinciding nodes with the leading $d_{x^2-y^2}$ eigenfunction.  However, this state is  inconsistent with 
the recent ultrasound measurements~\cite{ghosh2020thermodynamic,benhabib2020jump}, which find a discontinuity at $T_c$ of the elastic constant $c_{66}$ associated with shear B$_{2g}$ strain. Explaining this result requires two irreducible representations whose product transforms as $d_{xy}$\cite{ghosh2020thermodynamic}. 

The new experimental developments have inspired considerable theoretical activity\cite{Roising2019,suh2019,kivelson2020proposal,Gingras18,Huang18,Huang19,Wang19,Kaba19,Acharya19,Ramires2019,WangKallin20,Scaffidi2020,Wagner2020,Leggett2020}. Recently it was proposed that an accidental  two-component TRSB  state  of the form $d_{x^2-y^2}+ig_{xy(x^2-y^2)}$ could be consistent with the bulk of currently available experimental data\cite{kivelson2020proposal}. Both $d_{x^2-y^2}$ and $g_{xy(x^2-y^2)}$ phases were found to be dominant and
nearly degenerate in studies of the one-band Hubbard model including nearest-neighbor (NN) Coulomb interactions\cite{Raghu2012}, and therefore appear somewhat ``natural" as a candidate for \sruo.  Of course the electronic structure of \sruo ~is considerably more involved than the models considered in Ref. \onlinecite{Raghu2012}, so it is important to investigate the effects of longer-range Coulomb interactions for \sruo ~in the spin-fluctuation framework. As we show, NN Coulomb interaction is particularly important for \sruo~due to the near-degeneracy of symmetry-distinct superconducting phases of this material. 

Here, we perform a detailed theoretical study of the leading superconducting instabilities within a realistic electronic model for \sruo, and study the effects of both NN and next-nearest neighbor (NNN) Coulomb repulsion. The methodology includes as well the multi-orbital nature of this material, its significant SOC, and  onsite Hubbard-Hund interactions. We show that longer-range Coulomb interactions generally promote leading solutions of nodal $s'$- and $d_{xy}$-wave superconductivity over much of the phase diagram. This leads us to propose $s'+id_{xy}$-wave pairing as the most favorable superconducting candidate for \sruo. We analyze the properties of this TRSB even-parity state, and show that it exhibits a nodal spectrum, a superconducting spin susceptibility in agreement with neutron and NMR data\cite{Pustogow19,Ishida_correct,Chronister}, and symmetry properties consistent with recent ultrasound measurements\cite{ghosh2020thermodynamic,benhabib2020jump}. In  addition, for large Hund's coupling, longer-range Coulomb repulsion favors a TRSB mixed-parity $s'+ip$ solution.  We show that the leading state of this type also exhibits nodes, and could be consistent with Knight shift measurements. The $g$-wave state is also promoted by longer-range Coulomb repulsion, but remains non-competitive for \sruo. In addition, NN Coulomb repulsion disfavors the $d_{x^2-y^2}$ solution compared to most other states, implying that at least within the current framework, $d_{x^2-y^2}+ig_{xy(x^2-y^2)}$-wave superconductivity is not realized.

{\it Model and Method.} 
The non-interacting part of the Hamiltonian includes three orbitals $d_{xz}$, $d_{yz}$ and $d_{xy}$ with dispersions given by
$\xi_{xz}(\kv)=-2t_1\cos k_x -2t_2\cos k_y -\mu$,
$\xi_{yz}(\kv)=-2t_2\cos k_x -2t_1\cos k_y -\mu$, and
$\xi_{xy}(\kv)=-2t_3(\cos k_x +\cos k_y) 
-4t_4\cos k_x \cos k_y-2t_5(\cos 2k_x +\cos 2k_y) -\mu  
$ with $\{t_1,t_2,t_3,t_4,t_5,\mu\}=\{88,9,80,40,5,109\}$ meV~\cite{Zabolotnyy13,Cobo16}.
Atomic SOC, parametrized by $H_{SOC}=\lsoc \bf{L}\cdot\bf{S}$, gives rise to orbital mixing at the Fermi surface sheets, and a spin anisotropy of the magnetic susceptibility. We include SOC of $\lsoc \simeq 40$ meV ($\simeq 0.5t_1$).
Additionally, we incorporate both onsite and NN Coulomb interactions
\begin{eqnarray}
H_{nn}&=&\sum_{r_i,\delta,\mu,\nu,s,s'} \Big[W(\delta)\Big]^{\mu, s; \nu, s' }_{\nu,s'; \mu,s} n_{i,\mu,s} n_{i+\delta,\nu,s'},
\end{eqnarray}
where $n_{i,\mu,s}=c^\dagger_{i,\mu,s}c_{i,\mu,s}$ and $c^\dagger_{i,\mu,s}/c_{i,\mu,s}$ denotes creation/annihilation operator at site $i$ for an electron in orbital $\mu$ with spin $s$. In the following we use the joint index $\io=(\mu,s)$. The vector $\delta$ denotes the lattice vector between the sites;  $[W (\delta=0)]^{\io_1 \io_2}_{\io_3 \io_4}$ includes all onsite interactions and $[W (\delta)]^{\io_1 \io_2}_{\io_3 \io_4}$ with $\delta=\pm \hat x,\pm \hat y$ comprises NN Coulomb repulsion. Strong correlation effects important away from the Fermi surface are not included in our framework\cite{Mravlje11,Kim2018,Strand2019,Tamai2019,Zhang18,Kugler2020}, but are expected to leave the qualitative results of the spin-fluctuation pairing unchanged\cite{RomerPRL,RomerPRR2020}. We emphasize the importance of a sizable Hund's coupling, estimated as $J/U \simeq 0.1$\cite{Mravlje11,Vaugier12}.

\begin{figure*}[t]
    \centering
    \includegraphics[width=\linewidth]{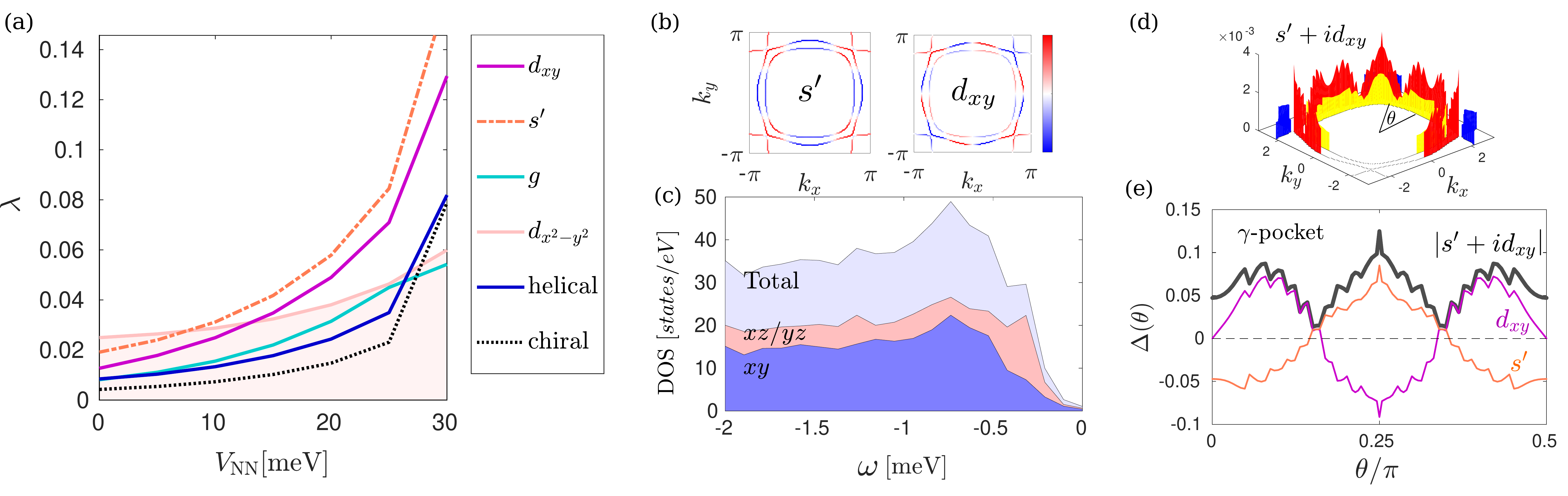}
\caption{(a) Superconducting instabilities as a function of nearest-neighbor repulsion $V_\mathrm{NN}$ for $U=100$ meV, $J/U=0.1$ for $\lsoc=35$ meV. The leading eigenvalue of each irreducible representation is shown.
(b) The two leading superconducting instabilities (for $V_{\rm NN}/U=0.2$ ), (c) orbital resolved density of states of the TRSB solution $s'+i d_{xy}$ for coupling parameters $V_\mathrm{NN}=20$ meV for $\lsoc=35$ meV, $U=100$ meV and $J/U=0.1$ for a maximum gap magnitude of $\Delta_0=1.5$ meV. (d) Spectral gap of the TRSB solution $s'+id_{xy}$, (e) angle dependent plot of the gap at the $\beta$-pocket as a function of angle in the first quadrant of the BZ.  
}
    \label{fig1}
\end{figure*}
In order to include both onsite and neighboring Coulomb interactions in the effective electron-electron interaction, it is useful to define a generalized bare susceptibility which includes $(\delta,\delta')$-dependent phase factors, see Eq.~(\ref{chi0}) below. We sum up all diagrams of bubble and ladder types to infinite order in the bare interactions. By this procedure, the effective pairing interaction consisting of spin- and charge-fluctuations is expressed in terms of the susceptibility in the multi-orbital RPA as stated in Eq.~(\ref{eq:Veffnn})
\begin{widetext}
\begin{eqnarray}
\Big[\chi_0(i\omega_n,\qv,\delta,\delta')\Big]^{\io_1 \io_2}_{\io_3 \io_4}
&=&\frac{1}{N}\int_0^\beta d\tau e^{i\omega_n \tau} \sum_{\kv} e^{i\kv(\delta-\delta')}
\langle c_{\kv,\io_2 }(\tau) c_{\kv,\io_3}^\dagger  \rangle_0  
\langle c_{\kv-\qv,\io_1 }^\dagger(\tau) c_{\kv-\qv,\io_4} \rangle_0,  \label{chi0} \\
\Big[V(\kv,\kv')\Big]^{\io_1,\io_2}_{\io_3,\io_4}&=&
\Big[W_0(\kv,\kv')\Big]^{\io_1,\io_2}_{\io_3,\io_4}+
\sum_{\delta,\delta'}e^{-i\kv\delta}e^{i\kv'\delta'}\Big[W\frac{1}{1-\chi_0W}\chi_0 W\Big]^{\io_1,\io_2}_{\io_3,\io_4}(\kv+\kv',\delta,\delta') \nonumber \\
&& \hspace{.5cm}-\sum_{\delta,\delta'}e^{-i\kv\delta}e^{-i\kv'\delta'}\Big[W\frac{1}{1-\chi_0W}\chi_0 W\Big]^{\io_1,\io_4}_{\io_3,\io_2}(\kv-\kv',\delta,\delta').
\label{eq:Veffnn}
\end{eqnarray}
\end{widetext}
The explicit form of the interaction matrices are stated in the Supplementary Material (SM).
In Eq.~(\ref{eq:Veffnn}), $\chi_0$ refers to the real part of the generalized susceptibility evaluated at zero energy. The resulting effective Hamiltonian 
\begin{equation}
 \hat{H}_{int}=\frac{1}{2}\!\sum_{ \kv,\kv' \{\tilde \mu\}}\!\!\Big[V(\kv,\kv')\Big]^{\io_1 , \io_2 }_{\io_3,\io_4 }  c_{\kv \io_1 }^\dagger  c_{-\kv \io_3 }^\dagger c_{-\kv' \io_2 } c_{\kv' \io_4 },
 \label{eq:Heff}
\end{equation}
is projected to band- and pseudospin-space, and the linearized BCS gap equation is subsequently solved
\begin{eqnarray}
  -\int_{FS} d \kv_f^\prime \frac{1}{v(\kv_f^\prime)} \Gamma_{l,l'}(\kv_f,\kv_f^\prime)\Delta_{l'}(\kv_f^\prime)=\lambda \Delta_l(\kv_f),
  \label{eq:LGE}
\end{eqnarray}
for the eigenvalue $\lambda$ and the gap function $\Delta_{l}(\kv_f)$ at wave vectors 
 $\kv_f$ on the Fermi surface. 
 The pairing kernel in band space is given by $\Gamma_{l,l'}(\kv_f,\kv_f^\prime)$, with spin information carried by the subscripts $l,l'=0,x,y,z$ which refer to the components of the ${\bf d({\bf k})}$-vector in pseudospin space\cite{SigristUeda}. Even-parity solutions appear in the $l=0$ channel.
 The Fermi surface is discretized by approximately 1000 wave vectors and $v(\kv_f)$ is the Fermi speed.
For further details we refer to the SM.

{\it Results.} As discussed in Refs.~\onlinecite{RomerPRL,Romer_strain2020,rmer2020fluctuationdriven}, spin-fluctuation-generated superconductivity for realistic models of \sruo ~including only local interactions tends to favor nodal $s'$, $d_{x^2-y^2}$, and helical pairing, depending on the amplitude of SOC and Hund's coupling $J$. In Fig.~\ref{fig1} we focus on a case where $d_{x^2-y^2}$ is leading in the case of onsite interactions only, and determine the evolution of the leading superconducting instabilities as a function of NN Coulomb repulsion, $V_\mathrm{NN}$. In the SM section we explore parameter space more broadly. As seen from  Fig.~\ref{fig1}, neither the $d_{x^2-y^2}$ nor $g_{xy(x^2-y^2)}$ wave solutions are promoted relative to the other channels by $V_\mathrm{NN}$.  To our knowledge the ratio $V_\mathrm{NN}/U$ has not been calculated by microscopic methods such as constrained RPA; however for cuprates with identical crystal structure, the corresponding ratio is close to 0.2\cite{Hirayama2018}. Importantly, for a substantial range of $V_\mathrm{NN}$ the two leading (nearly degenerate) instabilities exhibit $s'$- and $d_{xy}$-wave symmetries, as seen from  Fig.~\ref{fig1}(a). As shown in the SM section, this is a robust result relevant for a large region of parameter space: $V_\mathrm{NN}$ favors the $s'$ and $d_{xy}$ channels. Interestingly, all pairing channels are enhanced in absolute terms by $V_{\rm NN}$ (Fig. \ref{fig1}a). At first glance this may seem counter-intuitive; for example, an NN $d_{x^2-y^2}$ state should be suppressed by a NN repulsion.  Here, we find that the eigenvalue corresponding the pairing eigenfunction with this symmetry is weakly enhanced and evolves  to extend its range to avoid the NN interaction, as discussed in the SM.

The clear dominance of $s'$ and $d_{xy}$ channels suggests a new candidate state as a natural two-component TRSB superconducting phase stabilized by longer-range Coulomb repulsion: a $s'+id_{xy}$ condensate. Since it is comprised of A$_{1g}$ and B$_{2g}$ irreducible representations, the product clearly transforms as $d_{xy}$. Superconductivity of the form $s'+id_{xy}$ is therefore both consistent with a discontinuity at $T_c$ of the shear modulus $c_{66}$ associated with shear B$_{2g}$ strain, and the absence of any discontinuity associated with B$_{1g}$ strain\cite{ghosh2020thermodynamic,benhabib2020jump}. Below we further analyse the properties of this phase, and demonstrate that its symmetry, spectral and magnetic properties also agree with experiments.  We note that an $s'+id_{xy}$-like state with inter-orbital spin-triplet character generated by an entirely different mechanism (non-local SOC) was recently explored by Clepkens {\it et al.}\cite{Kee2020}.

Figures~\ref{fig1} (b)-(e) provide an overview of the spectroscopic properties of the $s'+id_{xy}$ gap structure. As seen from Fig.~\ref{fig1}(b), the constituents, $s'$ and $d_{xy}$, both feature a nodal spectrum: whereas $d_{xy}$ exhibits the required symmetry-imposed nodes along $k_x=0$ and $k_y=0$, both $s'$ and $d_{xy}$ contain additional nesting-driven nodes slightly off the $k_x=\pm k_y$ diagonal directions. These nodes are driven by the joint requirement of 1) even-parity, and 2) gap sign changes on FS sheets connected by the leading $\Qv_1$ and $\Qv_3$ nesting vectors (see SM for further details and the robustness of these nodes). The nodes are positioned at points on the FS where the orbital content changes its main character because the pairing interaction is weakest at these locations. Because these nodes coincide for both components, the ground state $s'+id_{xy}$ is also nodal, as shown in Fig.~\ref{fig1}(d,e). The resulting density of states (DOS) is displayed in Fig.~\ref{fig1}(c). In summary, both $s'$ and $d_{xy}$ contain vertical line nodes, which the lower-temperature TRSB phase $s'+id_{xy}$ inherits.

We note that while the existence of nodes seems to be well-established, their location on the Fermi surface is controversial. For example, Deguchi {\it et al.}\cite{Deguchi04} reported vertical nodes in the (100) direction from field angle-dependent specific heat, Kittaka {\it et al.}\cite{Kittaka2018} proposed the existence of horizontal nodes at nonzero $k_z$ based on a 3D version of the same measurement, while recent STM quasiparticle interference data was reported to be consistent  with nodes in the (110) direction\cite{ghosh2020thermodynamic}. Here, we predict vertical nodes located close to, but not identical with, the (110) direction.

\begin{figure}[t]
    \centering
    \includegraphics[width=\linewidth]{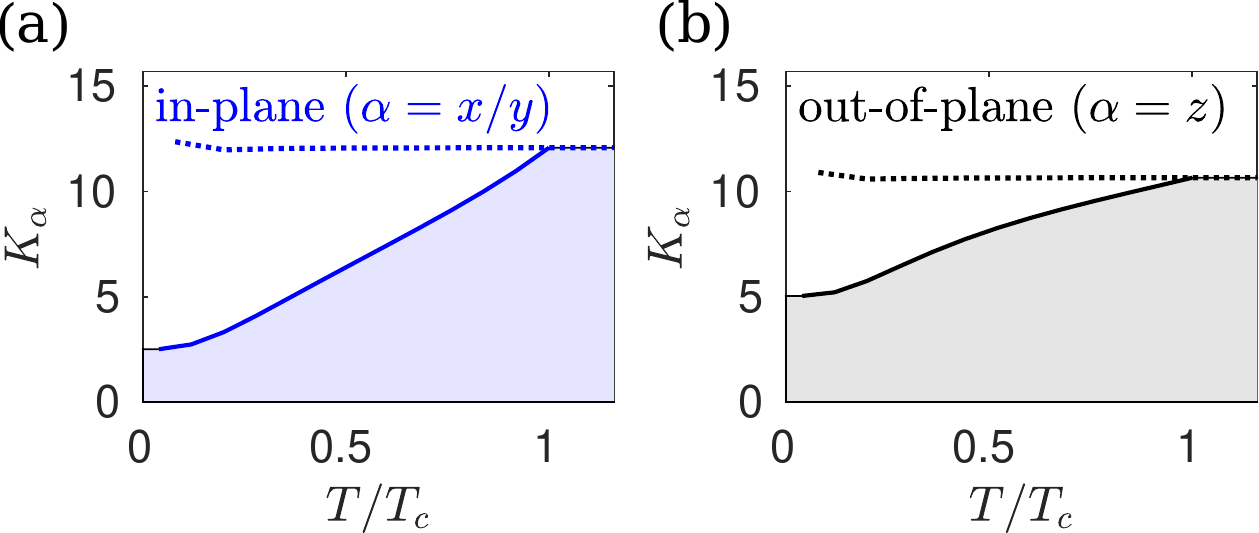}
    \caption{(a) In-plane  and (b) out-of-plane Knight shift obtained from Eq.~(\ref{eq:Knight}) for the solution $s'+id_{xy}$ with $U=170$ meV and $J/U=0.1$ ($V_\mathrm{NN}/U=0.2$). Maximal gap at $T=0$ is set to $\Delta_0=1.5$ meV. Solid (dashed) lines refer to the Knight shift in the superconducting (normal) state.}
    \label{fig:NMR}
\end{figure}

Turning next to a discussion of NMR Knight shift data, the  even-parity character of the $s'+id_{xy}$ state is not sufficient to guarantee agreement with the experimental data, simply because of the  significant SOC which mixes spin-singlet and triplet states. Therefore we perform a detailed microscopic calculation for $0<T<T_{c}$, assuming BCS-like temperature dependence of all gaps.
In the linear response regime, the Knight shift for an external magnetic field along $ \alpha \in\{x,y,z\}$ is obtained from the static spin-resolved susceptibility
\begin{equation}
K_\alpha =\frac{1}{4} \sum_{\mu,\nu,\{s\}} \spin_{s_1,s_2}^\alpha \spin_{s_3,s_4}^\alpha [\chi_{\rm sc}]^{\mu, s_1; \mu,s_2}_{\nu, s_3;\nu s_4},
\label{eq:Knight}
\end{equation}
where $\spin^\alpha$ denotes the Pauli matrices, and 
\begin{eqnarray}
\Big[\chi_{\rm sc}\Big]^{\io_1,\io_2}_{\io_3,\io_4}&=&  \Big[\frac{1}{1-\chi_{\rm sc,0}W}\chi_{\rm sc,0} \Big]^{\io_1,\io_2}_{\io_3,\io_4}(\qv,i\omega_n,\delta,\delta'),
\end{eqnarray}
is the RPA spin susceptibility in the superconducting state evaluated at $\qv=(0,0)$, $i\omega_n=0$ and $\delta=\delta'=0$. In Fig.~\ref{fig:NMR} we show the temperature evolution of the Knight shift for both in-plane and out-of-plane fields where in particular the in-plane field gives rise to a suppression of $K_{x,y}$ at low temperatures, consistent with the recent report by Chronister {\it et al}\cite{Chronister}. The magnitude of the suppression depends on the size of SOC and interaction parameters, as discussed in detail in SM.

{\it Discussion and conclusions.} We have developed a general theoretical framework for superconducting pairing within the spin-fluctuation scenario in the presence of SOC and longer-range Coulomb repulsion, and applied this to \sruo. Specifically,  we presented the consequences of $V_\mathrm{NN}$ in the pairing kernel, while effects of $V_\mathrm{NNN}$ are discussed in the SM. The hierarchy of the leading symmetry solutions changes as a function of $V_\mathrm{NN}$, and a leading $s'+id_{xy}$ state is favored, and proposed here as leading candidate  ground state for \sruo. Within this scenario, the momentum structure of the spin susceptibility is crucially important. As in Refs. \onlinecite{RomerPRL,Romer_strain2020}, we have constrained the interaction parameter space by requiring that the magnetic susceptibility be consistent with neutron measurements. For the current proposal, the position of the nesting-driven nodes is tied to the presence of a prominent $\Qv_3$ peak (see SM) in the susceptibility, which has been observed by a time-of-flight neutron scattering experiment\cite{Iida11}, but is controversial since it has not been confirmed by triple axes neutron scattering. We predict that future neutron scattering measurements should detect an enhanced scattering cross section at this momentum transfer. Another prediction is seen from Fig.~\ref{fig:NMR} in terms of the Knight shift evolution in the presence of out-of-plane fields. Of course, the detailed gap structure and in particular the presence of nesting-enforced nodes shown in Fig.~\ref{fig1} constitutes another property of the current theory, in principle verifiable by high-resolution momentum-dependent quasiparticle probes.   

While the $d_{x^2-y^2}+ig_{xy(x^2-y^2)}$ state is not favored in our calculations, other competing states are not completely ruled out by the spin-fluctuation analysis. The helical triplet state, favored for larger $J$\cite{RomerPRL} and sizable SOC, displays a substantial Knight shift suppression for in-plane fields and
is relatively unaffected by longer-range Coulomb interactions. As elaborated in the SM section, a (smaller) region of phase space would naturally support helical pairing, and a TRSB mixed parity $s'+i p$ state\cite{Eschrig2001,Scaffidi2020}. 
Surprisingly, this state also exhibits a nodal gap structure induced by $V_{\rm NN}$ and a Knight shift suppression consistent with NMR experiments (see SM). On the other hand, its agreement with the recent ultrasound experiments remains questionable at present.

{\it Acknowledgements.} We thank I. Eremin, A. Kreisel and S. Mukherjee for insightful conversations. A.T.R. and B.M.A. acknowledge support from the Carlsberg Foundation. B. M. A. acknowledges support from the Independent Research Fund Denmark grant number 8021-00047B. P. J. H. was supported by the U.S. Department of Energy under Grant No. DE-FG02-05ER46236. 
 \bibliography{bibliography_LongerCoulomb}
\pagebreak[2]
\begin{center}
\section{Supplementary Material}
 \end{center}
This Supplementary Material (SM) provides additional details about the leading and sub-leading instabilities of the superconducting order in \sruo, when longer-range Coulomb interaction is included. The derivation of the effective pairing interaction from spin- and charge-fluctuations in the presence of spin-orbit coupling (SOC) and longer-range Coulomb interaction is outlined.

Furthermore, we provide information on the spin susceptibility and additional details about the superconducting order parameters obtained from spin-fluctuation mediated pairing and the detailed dependence on interaction parameters.

\section{Pairing with longer-range Coulomb interaction}
 We will develop a formalism that takes into account onsite and nearest-neighbor (and next-nearest neighbor) interactions on an equal footing in the effective pairing interaction within the formalism of spin-fluctuation mediated superconductivity. In addition, this SM section contains a survey of the parameter dependence of our results presented in the main manuscript.

\begin{figure}[b]
\includegraphics{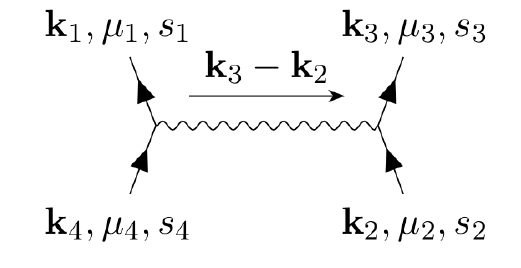}
\caption{Bare interaction diagram.}
\label{fig:bareV}
\end{figure}

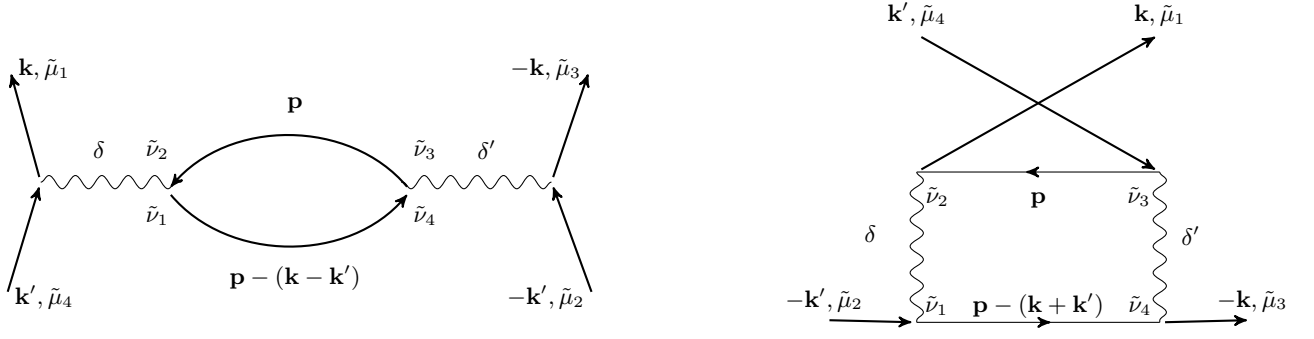
\begin{figure*}[t]
\centering
\begin{tikzpicture}[node distance=1cm, auto,]
 \node[] (markets) {};
 \node[right=3cm of markets] (market) {}
 (market.south) edge[pil, bend right=50] (markets.south);
 \node[right=2cm of markets] (0) {} 
 (markets.south) edge[pil, bend right=50] (market.south);
\node[left=4.8cm of market] (in) {};
\node[left=2.7cm of market] (nu2) {}
(in) edge[photon] (nu2);
\node[above=.2cm of nu2] (nu2wi) {};
\node[left=.9cm of nu2wi] {$\delta$};
\node[left=.05cm of nu2wi] (nu2w) {$\tilde \nu_2$};
\node[below=0.4cm of nu2w] (nu1w) {$\tilde \nu_1$};
\node[left=.01cm of market] (nu3) {};
\node[right=2.cm of nu3] (in2) {}
(in2) edge[photon] (nu3);
\node[above=.2cm of nu3] (nu3wi) {};
\node[right=.9cm of nu3wi]  {$\delta'$};
\node[right=.05cm of nu3wi] (nu3w) {$\tilde \nu_3$};
\node[below=0.4cm of nu3w] (nu3w) {$\tilde \nu_4$};
\node[above=.8cm of market](ppr) {};
\node[left=1.2cm of ppr](pp) {$\pv$};
\node[below=1.8 of pp](0) {$\pv-(\kv-\kv')$};
\node[left=3cm of pp] (mu1b) {};
\node[above=.1cm of mu1b] (mu1) {$\kv,\tilde\mu_1$}
(in.east) edge[pil] (mu1.west);
\node[below=2.5cm of mu1] (mu4) {$\kv',\tilde\mu_4$}
(mu4.west) edge[pil] (in.east);
\node[right=3cm of pp] (mu3b) {};
\node[above=.1cm of mu3b] (mu3) {$-\kv,\tilde\mu_3$}
(in2.west) edge[pil] (mu3.east);
\node[below=2.5cm of mu3] (mu2) {$-\kv',\tilde\mu_2$}
(mu2.east) edge[pil] (in2.west);

 \node[right=6.5cm of market] (ladderd) {};
 \node[below=.75cm of ladderd] (ladder) {};
 \node[above=1cm of ladder] (l1) {};
 \node[right=3cm of l1] (l4) {}
 (l4.south) edge[electron] (l1.south);
 \node[below=2cm of l4] (l3) {};
 \node[below=2cm of l1] (l2) {}
 (l2.north) edge[electron] (l3.north)
 (l1.south) edge[photon] (l2.north)
 (l4.south) edge[photon] (l3.north);
  \node (pp) at ($(l1)!0.5!(l4)$) {};
  \node[below=.9mm of pp](){$\tilde \nu_2\quad\qquad~\pv~\quad\qquad\tilde \nu_3$};
  \node[below=1.5cm of pp](){$\tilde \nu_1 \quad\pv-(\kv+\kv') \quad\tilde\nu_4$};
 \node[right=1.8cm of pp](ppr){};
 \node[left=2.cm of pp](ppl){};
 \node[below=0.6cm of ppr]{$\delta'$};
 \node[below=0.6cm of ppl]{$\delta$};
 
  \node[above=1.6cm of l1](l4in){$\kv',\tilde \mu_4$}
(l4in.south) edge[pil] (l4.south);
  \node[above=1.6cm of l4](l1out){$\kv,\tilde \mu_1$}
(l1.south) edge[pil] (l1out.south);

  \node[left=1cm of l2] (l2in) {};
  \node[above=.03cm of l2in] (l2inb) {$-\kv',\tilde \mu_2$}
(l2inb.south) edge[pil] (l2.north);
  \node[right=1cm of l3] (l3out) {};
  \node[above=.03cm of l3out] (l3outb) {$-\kv,\tilde \mu_3$}
(l3.north) edge[pil] (l3outb.south);

\end{tikzpicture}
\caption{Second order bubble and ladder diagrams. Note that each interaction line $V$ carries four joint indices $\tilde \mu=(\mu,s)$ for orbital and electronic spin and a $\delta$-vector labelling the range of the bare Coulomb interaction.}
\label{fig:bubbleladder}
\end{figure*}

We define the bare Coulomb repulsion Hamiltonian (with onsite and longer range interactions) as
\begin{eqnarray}
H&=&\sum_{r_i,\delta,\mu,\nu,s,s'} W(\delta) n_{i,\mu,s} n_{i+\delta,\nu,s'}\nonumber \\
&=&\sum_{r_i,\delta,\{\io\}} \Big[ W(\delta)\Big]^{\io_1,\io_2}_{\io_3,\io_4} c^\dagger_{i\io_1}  c^\dagger_{i+\delta,\io_3} c_{i+\delta,\io_2} c_{i,\io_4},
\end{eqnarray}
where $\delta$ defines the lattice vector between sites and $W(\delta)$ is the strength of the bare Coulomb repulsion between those sites.
In the second line we have introduced generalized indices for the longer-range Coulomb interaction and collected orbital and spin index in one common index $\io=(\mu,s)$. \\
\noindent By Fourier transformation of the bare interaction we get:
\begin{eqnarray}
H&=&\sum_{\{\kv \},\{\io\},\delta}
\Big[W(\delta)\Big]^{\io_1,\io_2}_{\io_3,\io_4} e^{i\delta(\kv_3-\kv_2)}
c_{\kv_1\io_1}^\dagger c_{\kv_3 \io_3}^\dagger c_{\kv_2 \io_2}c_{\kv_4\io_4}, \nonumber \\
\label{eq:Hbare}
\end{eqnarray}
\noindent i.e. the bare interaction depends explicitly on the momentum transfer and  $\kv_3-\kv_2=\kv_4-\kv_1$ due to momentum conservation.\\
First, we consider the summation of bubble diagrams. The second order bubble diagram is depicted in Fig.~\ref{fig:bubbleladder} (left). Since the transferred momentum $\kv_3-\kv_2=\kv'-\kv$ is independent of the internal momentum index $\pv$ of the bare susceptibility, it is possible to directly perform the $\delta$-sum in Eq.~(\ref{eq:Hbare}). Including onsite and nearest-neighbor interactions, the bare interaction elements are expressed as a function of $\qv=\kv-\kv'$ in the following way: \\
{\small
\underline{Same orbitals}
\begin{eqnarray}
&&\Big[W(\qv)\Big]^{\mu s \mu s}_{\mu  s' \mu  s'}=\Big\{\begin{array}{cl}
     -U -2V[\cos(q_x)+\cos(q_y)] & s=-s' \\
     -2V[\cos(q_x)+\cos(q_y)] & s=s'
\end{array} \nonumber
\end{eqnarray}
\underline{Different orbitals} ($\mu\neq \nu$)
\begin{eqnarray}
&&
\quad \Big[W(\qv)\Big]^{\nu s \nu s}_{\mu  s' \mu  s'}=\Big\{\begin{array}{cl}
     -U' -2V[\cos(q_x)+\cos(q_y)] & s=-s' \\
     -U'+J-2V[\cos(q_x)+\cos(q_y)] & s=s' 
\end{array} \nonumber
\end{eqnarray} 
 \begin{eqnarray} 
\quad \Big[U\Big]^{\mu s\nu s}_{\mu \overline s\nu \overline s}=-J',\quad  
\Big[U\Big]^{\mu s\nu s}_{\nu \overline s\mu \overline s}=-J.
 \label{eq:BareBubble}
\end{eqnarray}
}
The second order bubble diagram is evaluated as
\begin{eqnarray}
V_{(bub)}^{(2)}&=&-
[W(\kv-\kv')]^{\io_1\io_4}_{\inu_1\inu_2}
[\chi_0(\kv-\kv')]^{\inu_1\inu_2}_{\inu_3\inu_4}
[W(\kv-\kv')]^{\inu_3\inu_4}_{\io_3\io_2} \nonumber \\
\label{eq:Bub2}
\end{eqnarray}
with the generalized susceptibility as defined below in Eq.~(\ref{eq:BareSus}) evaluated with $\delta=\delta'=0$. Summation in $\inu$ indices in Eq.~(\ref{eq:Bub2}) is implicit.
The summation of all bubble diagrams gives
\begin{eqnarray}
&&V_{bub}(\kv,\kv')=-[W(\kv-\kv')\chi_{\rm RPA}(\kv-\kv')W(\kv-\kv')]^{\io_1, \io_4}_{\io_3, \io_2} \nonumber \\
&&\chi_{\rm RPA}(\kv-\kv')=[1-\chi_0(\kv-\kv') W(\kv-\kv')]^{-1}\chi_0(\kv-\kv')\nonumber \\
\end{eqnarray}
where matrix multiplication in orbital and spin indices is implicit and the susceptibility is evaluated for $\delta=\delta'=0$. Note that the sign convention of Eq.~(\ref{eq:BareBubble}) ensures an alternating sign between the effective interaction terms containing an even and odd number of bubbles, respectively.

In order to proceed with the ladder diagrams we 
consider the bare onsite interactions 
{\small
\begin{eqnarray}
 &&\Big[W(0)\Big]^{\mu s \mu  \overline s}_{\mu  \overline s \mu s}=U \qquad \Big[W(0)\Big]^{\nu s \mu  \overline s}_{\mu  \overline s\nu  s}=U' \qquad \Big[W(0)\Big]^{\mu s\nu \overline s}_{\mu \overline s\nu s}=J' \nonumber \\
&& \Big[W(0)\Big]^{\mu s\mu \overline s}_{\nu \overline s\nu s}=J \qquad \Big[W(0)\Big]^{\mu s\nu s}_{\nu s\mu s}=U'-J, \nonumber \\
 \label{eq:U}
\end{eqnarray}}
which are symmetry-related to the onsite interactions of Eq.~(\ref{eq:BareBubble}) and contains intra- and interorbital interactions. As above, it is understood that $\mu \neq \nu$. 
The longer-range interactions contain an explicit reference to momentum. 

In the case of ladder diagrams, the $\delta$-sum of Eq.~(\ref{eq:Hbare}) cannot be performed since the internal momentum label of the susceptibility enters the interaction vertex directly. Therefore, we keep the definition general and define for intraorbital interactions
\begin{eqnarray}
&&\Big[W(\kv_1,\kv_4,\delta)\Big]^{\mu, s ;\mu, s}_{\mu , s; \mu ,s}=  V(\delta) e^{i\delta(\kv_4-\kv_1)}, \nonumber \\
&&\Big[W(\kv_1,\kv_4,\delta)\Big]^{\mu ,s ;\mu ,\overline s}_{\mu , \overline s; \mu, s}=V(\delta)e^{i\delta(\kv_4-\kv_1)},
\label{eq:Vt1}
\end{eqnarray}
and interorbital interactions ($\mu \neq \nu$)
\begin{eqnarray}
&&\Big[\tilde W(\kv_1,\kv_4,\delta)\Big]^{\mu, s ;\nu, s}_{\nu,  s; \mu, s}= V(\delta)e^{i\delta(\kv_4-\kv_1)},\nonumber \\
&&\Big[\tilde W(\kv_1,\kv_4,\delta)\Big]^{\mu ,s ;\nu, \overline s}_{\nu,  \overline s; \mu, s}= V(\delta)e^{i\delta(\kv_4-\kv_1)}.
\label{eq:Vt2}
\end{eqnarray}
For simplicity we set the intra- and interorbital interactions to be the same. For nearest-neighbor Coulomb repulsion we have $V(\delta)=V_\textrm{NN}$  with $\delta \in \{\hat x,-\hat x,\hat y,-\hat y\}$ and next-nearest-neighbor Coulomb repulsion, $V(\delta)=V_\textrm{NNN}$ with $\delta \in \{\hat x+\hat y,-\hat x+\hat y,-\hat x -\hat y,\hat x-\hat y\}$.

In the derivation of the effective pairing interaction for longer-range interactions we transfer the $\delta$- and $\kv$-dependence of the bare interaction line (see Fig.~\ref{fig:bareV}) to the susceptibility. Therefore it proves useful to define a generalized bare susceptibility including a phase which depends on the vectors $\delta,\delta'$. These vectors belong to the set $\{0,\hat x,\hat y,-\hat x,-\hat y,\hat x+\hat y, -\hat x+\hat y,\hat x-\hat y,\hat x-\hat y\}$. 
In this way, the second order ladder diagram, for example, is evaluated to yield
\begin{widetext}
\begin{eqnarray}
V_{lad}^{(2)}(\kv,\kv')&=&\sum_{\delta,\delta'}[V(\delta)]^{\io_1 \io_2}_{\inu_1 \inu_2}e^{-i\delta\kv}e^{i\delta'\kv'}\underbrace{\sum_\pv e^{i\pv(\delta-\delta')} \langle c_{\pv \inu_2}(\tau)c^\dagger_{\pv \inu_3}\rangle_0 \langle c^\dagger_{\pv-(\kv+\kv') \inu_1}(\tau)c_{\pv-(\kv+\kv') \inu_4}\rangle_0} _{[\chi_0(\kv+\kv',\delta,\delta')]^{\inu_1 \inu_2}_{\inu_3 \inu_4}}
[V(\delta')]^{\inu_3 \inu_4}_{\io_3 \io_4},
\end{eqnarray}
where we have defined the generalized susceptibility which includes the real-space lattice vectors $\delta,\delta'$ as
\begin{eqnarray}
\Big[\chi_0(i\omega_n,\qv,\delta,\delta')\Big]^{\io_1 \io_2}_{\io_3 \io_4}&=&\frac{1}{N}\int_0^\beta d\tau e^{i\omega_n \tau} \sum_{\kv} e^{i\kv(\delta-\delta')}
\langle c_{\kv,\io_2 }(\tau) c_{\kv,\io_3}^\dagger  \rangle_0  
\langle c_{\kv-\qv,\io_1 }^\dagger(\tau) c_{\kv-\qv,\io_4} \rangle_0  \nonumber \\
&=&-\frac{1}{N}\sum_{\kv} e^{i\kv(\delta-\delta')}\sum_{n_1,n_2}[M_{n_1,n_2}(\kv,\qv)]^{\mu_1\spin_1,\mu_2\spin_2}_{\mu_3\spin_3,\mu_4\spin_4}
 \frac{f(\xi_{\kv-\qv, n_1 ,\spin_1})-f(\xi_{\kv ,n_2 ,\spin_2})}{i\omega_n+\xi_{\kv-\qv, n_1, \spin_1}-\xi_{\kv, n_2 ,\spin_2}},
\label{eq:BareSus}
 \end{eqnarray}
with
\begin{eqnarray}
 [M_{n_1,n_2}(\kv,\qv)]^{\mu_1\spin_1,\mu_2\spin_2}_{\mu_3\spin_3,\mu_4\spin_4}=[u_{n_1\spin_1}^{\mu_1 s_1}(\kv-\qv)]^*  [u_{n_2\spin_3}^{\mu_3 s_3} (\kv)]^* u_{n_2\spin_2}^{\mu_2 s_2} (\kv)u_{n_1\spin_4}^{\mu_4 s_4} (\kv-\qv).
\end{eqnarray}
Here $u_{n\spin}^{\mu s}(\kv)$ is the eigenvector of the transformation from orbital and electronic spin basis $(\mu,s)$ to band and pseudo-spin basis $(n,\sigma)$.
The generalized susceptibility becomes a $5\cdot 36 \times 5\cdot 36$ matrix due to the phases $e^{i\kv (\delta-\delta')}$ (when  next-nearest neighbors are included as well, it becomes a $9\cdot 36 \times 9\cdot 36$ matrix). 
We will explore superconductivity arising from spin fluctuations and the interaction Hamiltonian restricted to the Cooper channel:
\begin{eqnarray}
\hat H_{int}=\frac{1}{2}\sum_{ \kv,\kv' \{\tilde \mu\}}\Big[V(\kv,\kv')\Big]^{\io_1 , \io_2 }_{\io_3,\io_4 }  \quad c_{\kv \io_1 }^\dagger  c_{-\kv \io_3 }^\dagger c_{-\kv' \io_2 } c_{\kv' \io_4 },\nonumber \\
\label{eq:Hcooper}
\end{eqnarray}
includes the bare interaction 
as well as the effective interaction from bubble and ladder diagrams and vertex corrections consisting of admixtures of the bubble and ladder vertices. Thus we have
\begin{eqnarray}
\Big[V(\kv,\kv')\Big]^{\io_1,\io_2}_{\io_3,\io_4}&=&\Big[W_0(\kv,\kv')\Big]^{\io_1,\io_2}_{\io_3,\io_4}-\sum_{\delta,\delta'}e^{-i\kv\delta}e^{-i\kv'\delta'}\Big[W [1-\chi_0W]^{-1}\chi_0 W\Big]^{\io_1,\io_4}_{\io_3,\io_2}(\kv-\kv',\delta,\delta') \nonumber \\
&&+\sum_{\delta,\delta'}e^{-i\kv\delta}e^{i\kv'\delta'}\Big[W[1-\chi_0W]^{-1}\chi_0 W\Big]^{\io_1,\io_2}_{\io_3,\io_4}(\kv+\kv',\delta,\delta').\label{eq:Veffnn_Suppl}
\end{eqnarray}
The bare interaction is as stated in Eqs.~(\ref{eq:BareBubble}), (\ref{eq:U}), (\ref{eq:Vt1})  and (\ref{eq:Vt2}), i.e. for nearest neighbor repulsion it takes the form 
\begin{equation}
\hat H_{bare}=\sum_{\mu,\nu,s,s',\kv,\kv'}
2V_{\rm NN}[\cos(k_x-k_x')+\cos(k_y-k_y')]c_{\kv \mu s }^\dagger  c_{-\kv \nu s'}^\dagger c_{-\kv' \nu s' } c_{\kv' \mu s }
\end{equation}
with the same repulsion strength $V_{\rm NN}$ for intra- and interorbital as well as same and opposite spin interaction.

The RPA expression refers to the construction including the index $\delta$. Note that the matrices $W(\delta)$ are block-diagonal matrices of $36 \times 36$ blocks for each value of $\delta$.
\end{widetext}

The interaction vertex Eq.~(\ref{eq:Veffnn_Suppl}) enters the interaction Hamiltonian stated in Eq.~(\ref{eq:Hcooper}). 
In order to determine the superconducting instability at the Fermi surface, we project the interaction Hamiltonian to band (and pseudo-spin) space. In order to do so, we introduce 
fermion bilinear operators of the form
 \begin{eqnarray}
 \overline\Psi_l(n,\kv)&=&s_l\beta^\dagger_{\kv n\spin}[\Gamma_l]_{\spin\spin'}\beta^\dagger_{-\kv n'\spin'}\delta_{n,n'}, \nonumber \\
\Psi_l(n,\kv)&=&\beta_{-\kv n\spin}[\Gamma_l]_{\spin\spin'}\beta_{-\kv n'\spin'} \delta_{n,n'}.
\label{eq:Psi}
\end{eqnarray}
In the above definitions, $\spin$ denotes pseudo-spin, $s_{0,y}=-1$ and $s_{x,z}=+1$, and the $\Gamma_l$ matrices are constructed from the Pauli matrices $\sigma_l$ by
\begin{eqnarray}
 \Gamma_l&=&\frac{1}{\sqrt{2}}\sigma_l i \sigma_y .
\end{eqnarray}
This construction is similar to the well-known $\bf d$-vector, but refers to the {\it pseudo}-spin structure.
By use of the $\beta$-operators, we project the interaction Hamiltonian to band and pseudospin space:
\begin{eqnarray}
 \hat H_{int} 
&=&\frac{1}{2}\sum_{ \kv,\kv' \{n\} \{\spin\}}
 \sum_{l,l'} 
 \overline\Psi_{l}(n_1,\kv)
 \Gamma_{l,l'}(n_1 \kv;n_2 \kv')\Psi_{l'}(n_2,\kv') \nonumber \\
\end{eqnarray}
where the pairing kernel in band space is stated in Eq.~(\ref{eq:Vpseudo}) below, and projection into the four different pairing channels is carried out as stated in Eq.~(\ref{eq:lproj}). Here, $l=0$ is the even parity and pseudospin singlet channel, while $l=x,y$ are odd parity, pseudospin triplet channels with the pseudospins polarized out-of-plane and finally, $l=z$ is the odd parity, pseudospin triplet channel with the pseudospins polarized in plane
\begin{widetext}
\begin{equation}
[V(n_1,\kv;n_2,\kv')]^{\spin_1\spin_2}_{\spin_3\spin_4}=\sum_{\{\mu\}\{s\}} (u_{\mu_1 s_1}^{n_1 \spin_1}(\kv))^* (u_{\mu_3 s_3}^{n_1 \spin_3}(-\kv))^* 
\Big[V(\kv,\kv')\Big]^{\mu_1 s_1 , \mu_2 s_2 }_{\mu_3 s_3,\mu_4 s_4}  u_{\mu_2 s_2}^{n_2 \spin_2}(-\kv')  u_{\mu_4 s_4}^{n_2 \spin_4}(\kv'),\label{eq:Vpseudo}
\end{equation}
\begin{equation}
\Gamma_{l,l'}(n_1 \kv;n_2 \kv')=    \sum_{\{\tau\}} s_{l'}[\Gamma_l]_{\tau_3\tau_1}
[V(n_1,\kv;n_2,\kv')]^{\tau_1\tau_2}_{\tau_3\tau_4} 
[\Gamma_{l'}]_{\tau_4\tau_2}. \label{eq:lproj}
\end{equation}
This ends the discussion of spin-fluctuation mediated pairing in the presence of SOC and longer-range Coulomb interaction. Now we turn to a presentation of the spin susceptibility and superconducting instabilities arising from spin-fluctuation mediated pairing.
\end{widetext}

 \begin{figure}[t!]
    \includegraphics[width=\linewidth]{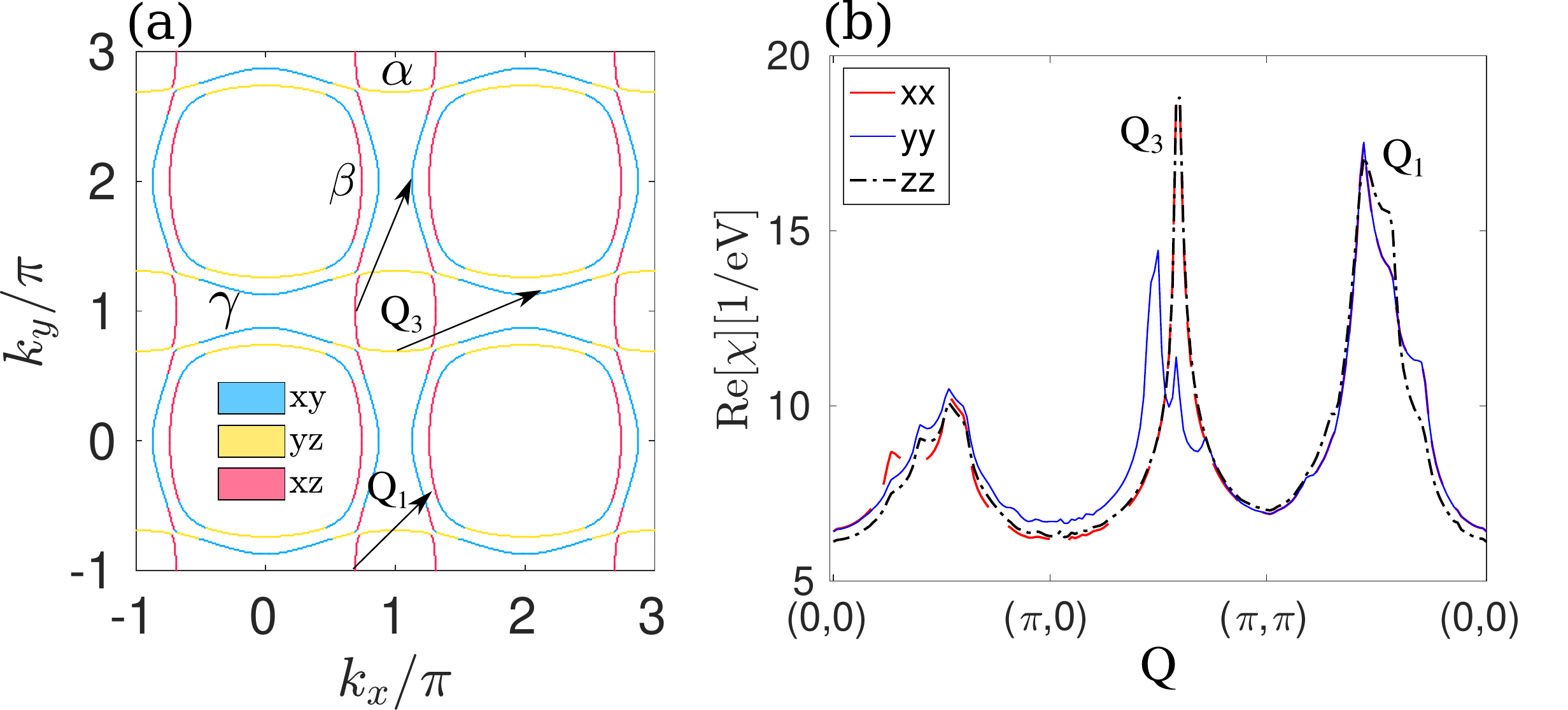}
    \caption{(a) Fermi surface for $\lsoc=35$ meV in the extended zone. Dominating orbital content  is shown by colors and the common nomenclature of the Fermi surface pockets $\alpha, \beta$ and $\gamma$ is indicated. The two nesting vectors $\Qv_1 \simeq (2 \pi/3,2 \pi/3) $ and $\Qv_3 \simeq (\pi/2,\pi) /(\pi,\pi/2)$ are depicted by black arrows. (b) Real part of the three RPA spin susceptibilities evaluated at zero energy: $\chi_{xx}$ (dashed red line), $\chi_{yy}$ (full blue line) and $\chi_{zz}$ (dashed-dotted black line). Interaction parameters $U=100$ meV, $J/U=0.1$ and $V_{NN}=10$ meV. The dominating peak is $\Qv_1$, which displays a spin anisotropy with a stronger out-of-plane component, which is visible from the energy-resolved susceptibility depicted in Fig.~\ref{fig:chiomega}.}
    \label{fig:FS_35meV}
\end{figure}

\section{Bands and spin susceptibility}
As stated in the main text, 
the non-interacting Hamiltonian includes the three orbitals $xz$, $yz$ and $xy$ with dispersions given by
$\xi_{xz}(\kv)=-2t_1\cos k_x -2t_2\cos k_y -\mu$,
$\xi_{yz}(\kv)=-2t_2\cos k_x -2t_1\cos k_y -\mu$, and
$\xi_{xy}(\kv)=-2t_3(\cos k_x +\cos k_y) 
-4t_4\cos k_x \cos k_y-2t_5(\cos 2k_x +\cos 2k_y) -\mu  
$ with $\{t_1,t_2,t_3,t_4,t_5,\mu\}=\{88,9,80,40,5,109\}$ meV.
The three orbitals $xz,yz$ and $xy$ are coupled by the SOC parametrized by $H_{SOC}=\lsoc \bf{L}\cdot\bf{S}$, and the splitting between the energy bands at the Fermi surface depends on the magnitude of the SOC.
We include a SOC of $\lsoc=35$ meV, as well as $\lsoc=45$ meV for comparison. Note that since we parametrize the bands by renormalized hopping constants this corresponds to a relatively 
large value of $\lsoc \simeq 0.5 t_1$, where $t_1$ is the largest hopping of the $xz (yz)$ band.
The resulting Fermi surface is shown in Fig.~\ref{fig:FS_35meV} (a), which also displays the dominating orbital character of each Fermi surface point. From this figure, the mixing of the orbital character in regions close to the Brillouin zone diagonals is clearly visible. 
We adopt the common nomenclature indicated in Fig.~\ref{fig:FS_35meV}(a): the 
$\alpha$-pocket is centered around $(\pi,\pi)$ and is primarily of $xz/yz$ orbital character, the $\beta$-pocket is centered around $(0,0)$ and is also of mainly $xz/yz$ orbital character except for the low-energy states in the vicinity of the zone diagonals. Finally the $\gamma$-pocket, centered at $(0,0)$, is mainly of $xy$ character with the exception of the low-energy states close to the zone diagonals which are primarily of $xz,yz$ orbital character. 
\begin{figure}[b!]
\includegraphics[width=0.75\linewidth]{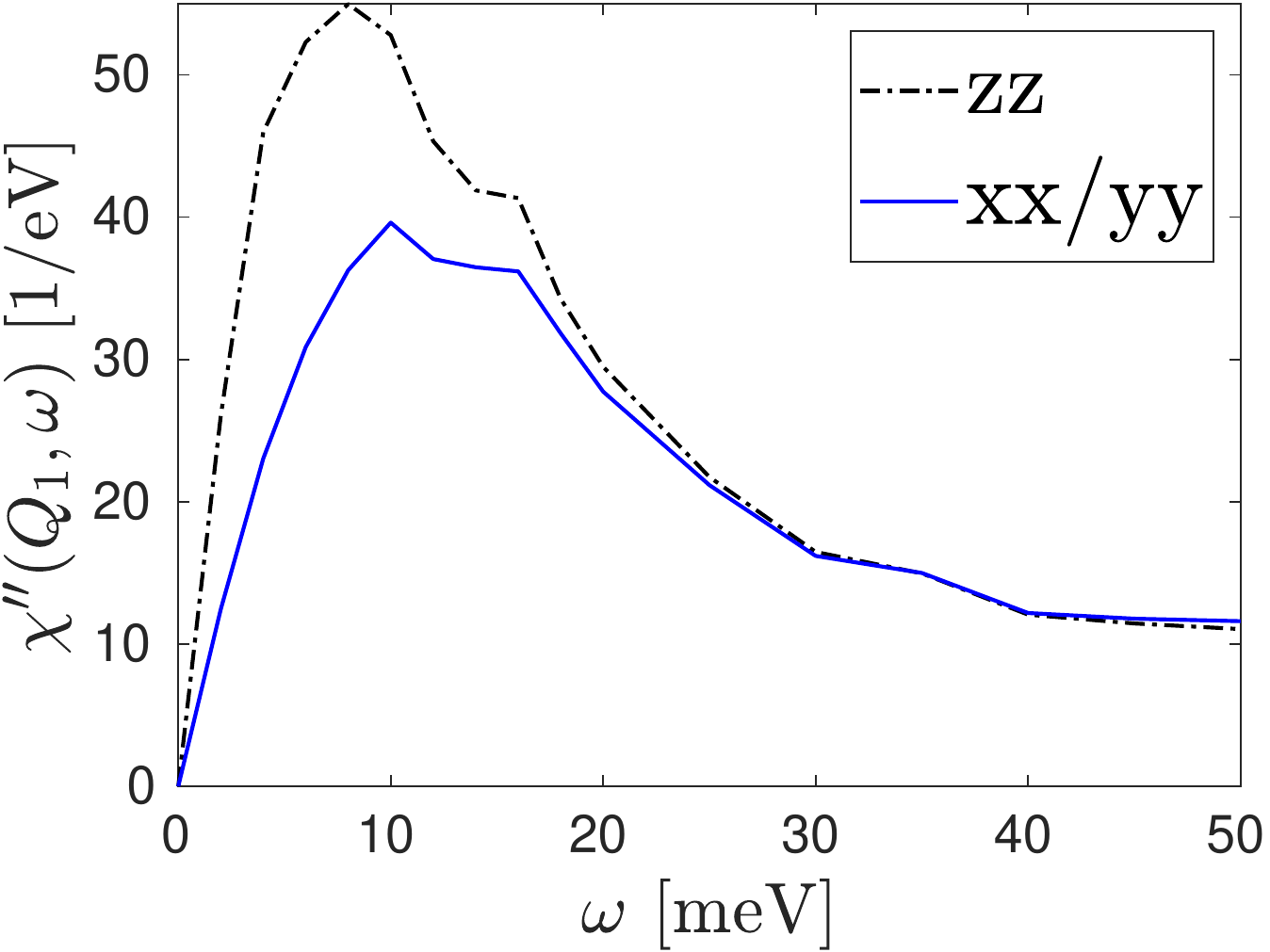}
\caption{Imaginary part of the RPA spin susceptibility at momentum $\Qv_1$ as a function of energy for $\lsoc=35$ meV and $U=170$ meV and $J/U=0.1$
 ($V_\textrm{NN}=0$). A pronounced enhancement of the out-of-plane ($zz$) spin susceptibility with respect to the in-plane components ($xx,yy$) is found at energies below 10 meV.}
    \label{fig:chiomega}
\end{figure}
\begin{figure*}
\includegraphics[width=\linewidth]{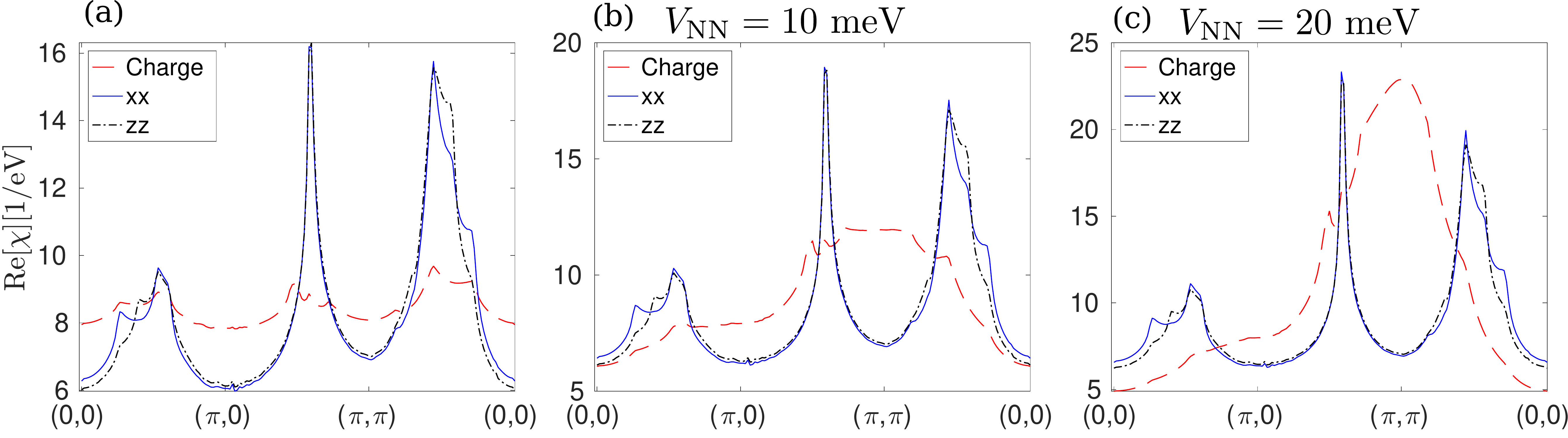}
\caption{The real part of the RPA spin ($zz$ and $xx$ components) and charge susceptibility evaluated a zero energy as a function of momentum $\Qv$ along the path $(0,0)-(\pi,0)-(\pi,\pi)-(0,0)$ for $\lsoc=35$ meV and $U=100$ meV and $J/U=0.1$ for
 $V_\textrm{NN}=0,10,20$ meV in (a,b,c). As the nearest neighbor interaction $V_\textrm{NN}$ is increased, the charge susceptibility becomes enhanced at $(\pi,\pi)$. Also the spin components of the RPA susceptibility show enhancements. }
    \label{fig:chiRPA_vnn}
\end{figure*}
The susceptibility including interactions in the RPA approximation is given by:
\begin{equation}
\Big[\chi_{\rm RPA}(\qv,i \omega_n,\delta,\delta')\Big]^{\io_1,\io_2}_{\io_3,\io_4}
= \Big[\frac{1}{1-\chi_0 W}\chi_0 \Big]^{\io_1,\io_2}_{\io_3,\io_4}(\qv,i\omega_n,\delta,\delta')
\end{equation}
where the $(\qv,i\omega_n)$-dependence is carried solely by $\chi_0$ and the summation in the indices $\mu,s,\delta$ for the RPA expression is implicit.
The physical susceptibility for the different spin channels $a,b \in\{x,y,z\}$ is obtained by a projection to spin channels and summation in orbital indices:
\begin{eqnarray}
 &&\Big[\chi^{\alpha \alpha'}(\qv,i\omega_n)\Big]= \frac{1}{N}\int_0^\beta d \tau e^{i \omega_n \tau} \sum_{\mu,\nu} \langle T_\tau S_{\mu}^\alpha (-\qv,\tau) S_{\nu}^{\alpha'} (\qv,0) \rangle \nonumber \\
&&= \frac{1}{4} \sum_{\mu,\nu} \spin_{s_1,s_2}^\alpha \spin_{s_3,s_4}^{\alpha'} [\chi (\qv,\delta=0,\delta'=0)]^{\mu, s_1; \mu,s_2}_{\nu, s_3;\nu, s_4}.\nonumber \\
 \end{eqnarray}
The RPA spin susceptibilities for the in-plane ($xx/yy$) and out-of-plane ($zz$) spin directions are depicted in Fig.~\ref{fig:FS_35meV} (b) and Fig.~\ref{fig:chiomega}.
The spin susceptibility has two prominent peak structures, which are  $\Qv_1 \simeq (2\pi/3,2\pi/3) $ and $\Qv_3 \simeq (\pi,\pi/2)/(\pi/2,\pi) $. We note that the exact position of $\Qv_1$ is sensitive to the Fermi surface nesting. For the given parameters, we get a rather broad $\Qv_1$ peak centered around $\Qv_1=(0.52,0.52)\pi$. This structure shows a spin anisotropy with a dominant out-of-plane spin component, an anisotropy which arises from the SOC and is enhanced by increasing onsite Coulomb interaction as well as the Hund's coupling interaction. In Fig.~\ref{fig:chiomega} we show the energy dependence of the inplane ($xx/yy$) and out-of-plane ($xx$) spin susceptibility  at wave vector $\Qv_1$.
 The out-of-plane ($zz$) spin susceptibility is roughly twice as large as the in-plane susceptibilities ($xx,yy$) at small energy transfers. The magnitude of this spin anisotropy depends sensitively on the interaction constants, which is taken to be relatively large in the case of Fig.
~\ref{fig:chiomega} with $\lsoc=35$ meV, $U=170$ meV and $J/U=0.1$. Smaller interaction strengths in general give a weaker anisotropy within RPA.
In neutron scattering experiments, the peak structure $\Qv_1$ has been reported to exhibit a large spin anisotropy of roughly a factor of two for intermediate energy transfers of $\hbar \omega =8$ meV~\cite{Braden04}.\\
In addition to the structure at $\Qv_1$, the peak at $\Qv_3$, which is indicated in Fig.~\ref{fig:FS_35meV}(b), has been reported in time-of-flight neutron measurements~\cite{Iida11}. However, it has not been confirmed by triple-axis spectroscopy.\\

 The charge susceptibility is calculated from the generalized susceptibility Eq.~(\ref{eq:BareSus}) by
 \begin{equation}
    \Big[\chi^{ch}(\qv,i\omega_n)\Big]
=\sum_{\mu,\nu,s_1,s_2} 
[\chi (\qv,\delta=0,\delta'=0)]^{\mu, s_1; \mu,s_1}_{\nu, s_2;\nu, s_2}.
 \end{equation}


\begin{figure}[b]
    \centering
        \includegraphics[width=\linewidth]{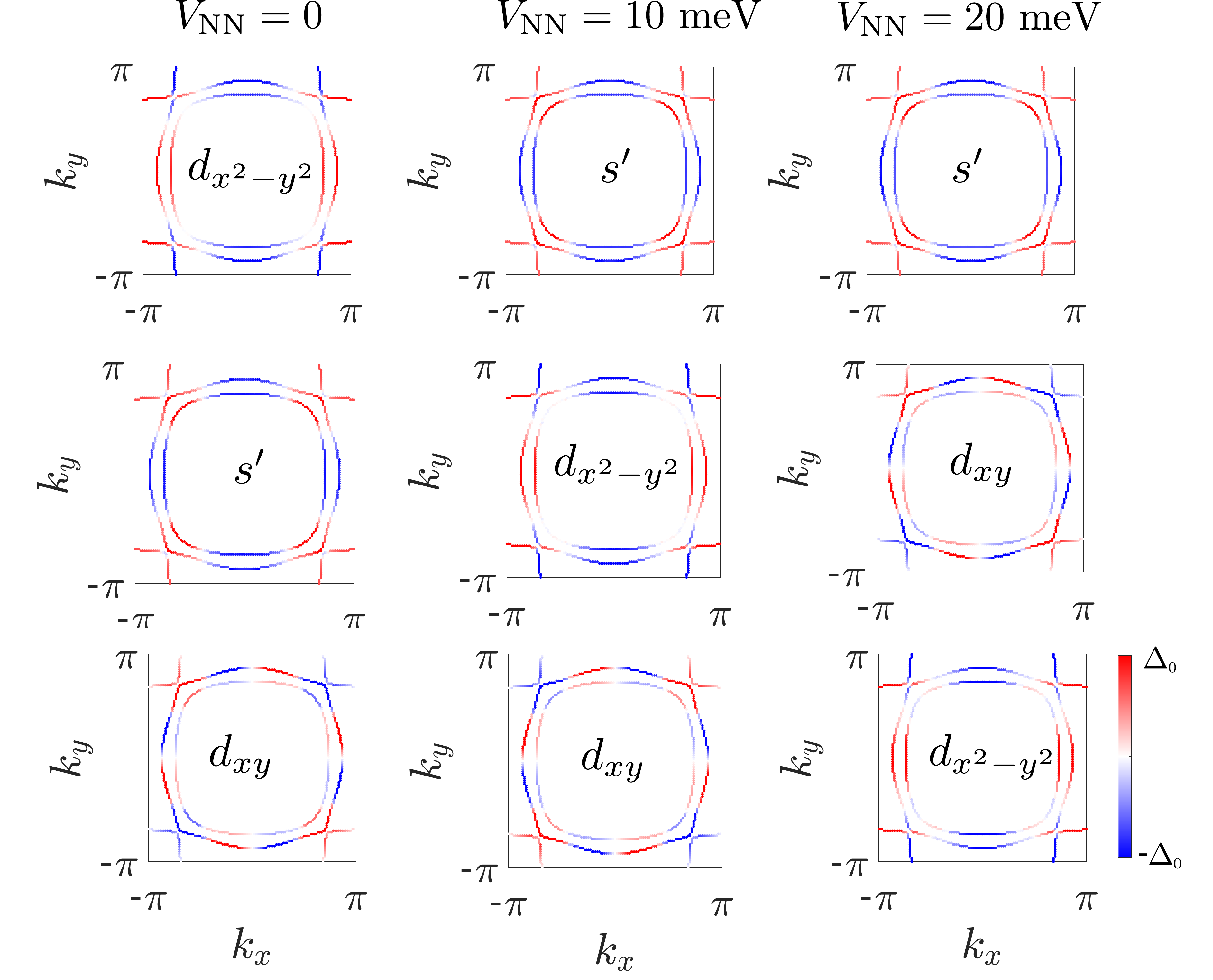}
        \caption{The three leading gap structures for nearest-neighbor Coulomb repulsion $V_\textrm{NN}=0,10,20$ meV for $U=100$ meV and $J/U=0.1$. }
        \label{fig:GapsThreeVnn}
\end{figure}
\begin{figure*}[t!]
\includegraphics[width=\linewidth]{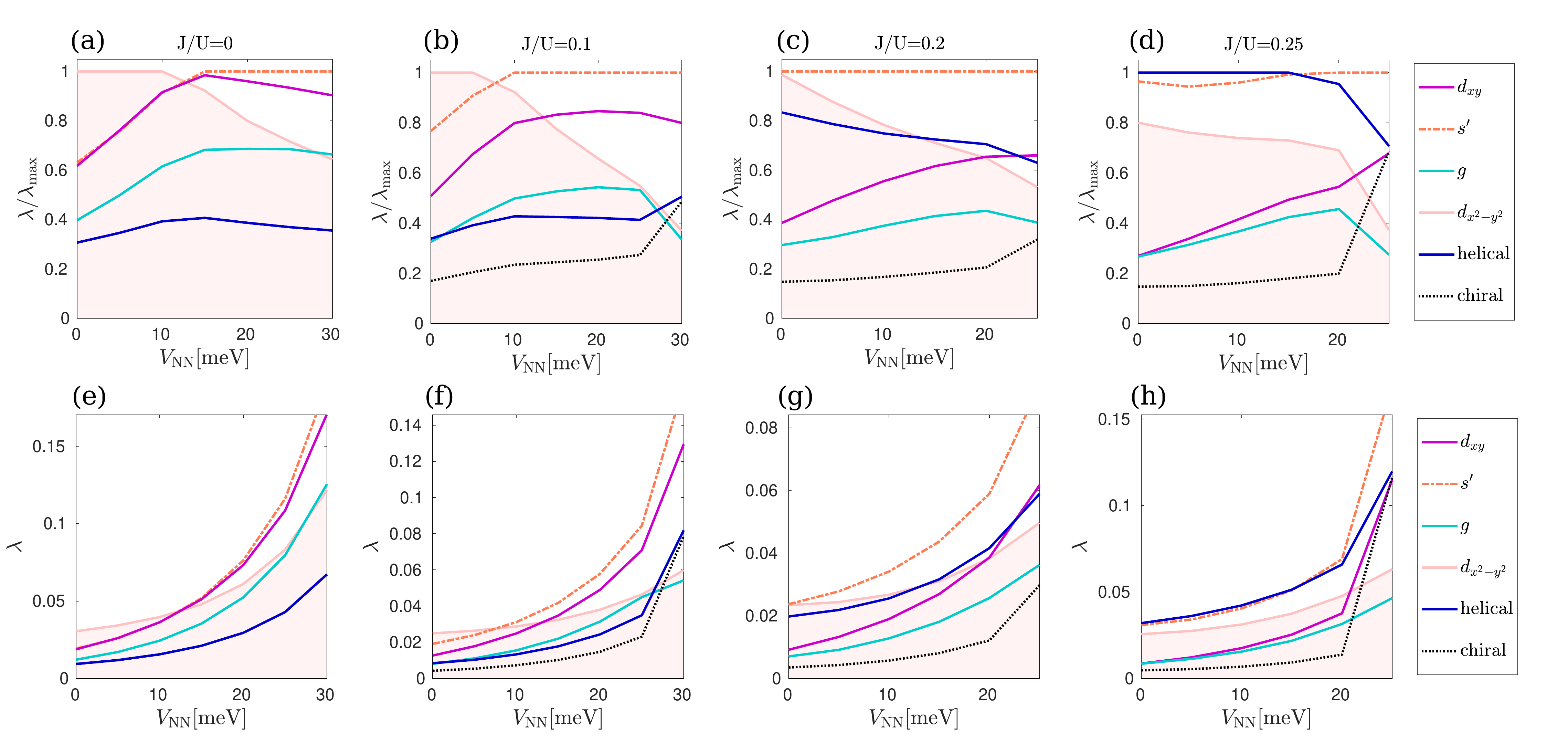}
\caption{Solutions to the linearized gap equation as a function of nearest-neighbor repulsion $V_\textrm{NN}$. Onsite Coulomb repulsion is $U=100$ meV and $\lsoc=35$ meV. Four different values of $J$ is chosen: (a) $J/U=0$ (note that in this case the helical and chiral solutions are degenerate and therefore the chiral solution is not visible), (b) $J/U=0.1$, and (c) $J/U=0.2$ and (d) $J/U=0.25$.
(a-d) For each value of $V_\textrm{NN}$, the eigenvalues of each irreducible representation is normalized to the maximal eigenvalue at that $V_\textrm{NN}$. The relative suppression of $d_{x^2-y^2}$ is clearly visible. (e-h) The evolution of the  eigenvalues of each symmetry channel on an absolute scale }
    \label{fig:DifferentHunds}
\end{figure*}
\begin{figure}[b!]
\includegraphics[width=0.75\linewidth]{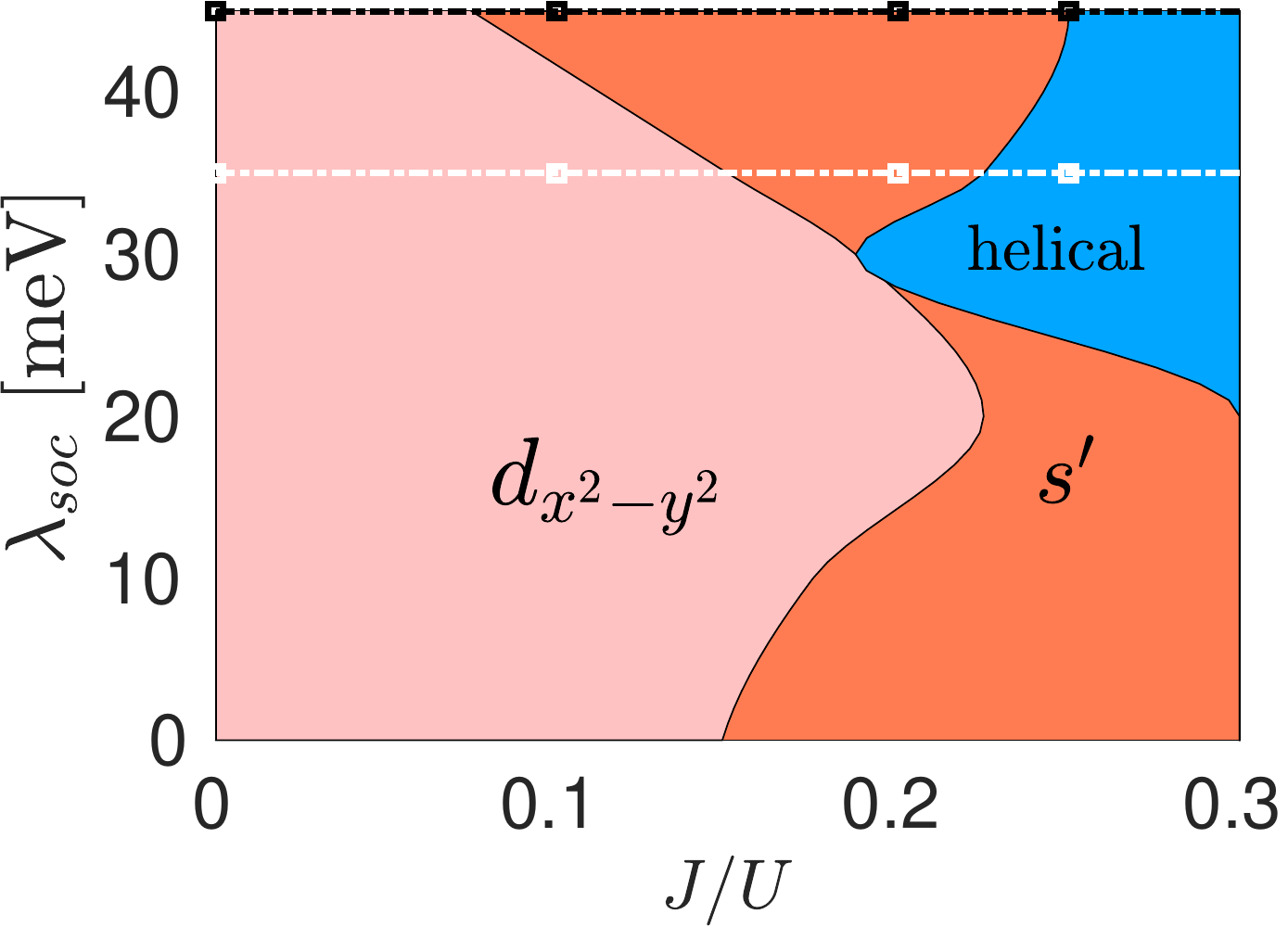}
\caption{Leading solution of the linearized gap equation as a function of Hund's coupling $J$ and spin-orbit coupling $\lsoc$ in the case of $U=120$ meV and zero 
nearest-neighbor repulsion $V_\textrm{NN}=0$. Reproduced from Ref.~\onlinecite{RomerPRL}. The white (black) squares indicate the starting points  for which we obtain the evolution of the superconducting instabilities as a function of $V_{\rm NN}$, see Figs.~\ref{fig:DifferentHunds} and \ref{fig:DifferentHunds_SOC45}, respectively.}
    \label{fig:Phasedia}
\end{figure}
\section{Structure of the superconducting instabilities}
The evolution of the different superconducting channels as a function of nearest-neighbor repulsion $V_\textrm{NN}$ was reported in the main text in Fig. 1 for Hund's coupling $J/U=0.1$. This value of Hund's coupling was estimated in Refs.~\onlinecite{Mravlje11,Vaugier12}. \\
In Fig.~\ref{fig:GapsThreeVnn} the leading and two subleading solutions are shown for three different values of $V_{\rm NN}$. For zero nearest-neighbor interaction the $d_{x^2-y^2}$ solution is leading and followed by nodal $s'$. The near-degeneracy between these two solutions has previously lead us to propose a gap structure of the form $s'+id_{x^2-y^2}$\cite{RomerPRL}. As seen from the first column in Fig.~\ref{fig:GapsThreeVnn}, the $d_{xy}$ solution appears as the third leading solution in the case of $V_{\rm NN}=0$. As $V_{\rm NN}$ is increased, the $d_{x^2-y^2}$ solution is suppressed compared to $s'$ and $d_{xy}$ and for  $V_{\rm NN}=20$ meV, the latter two are the leading solutions.\\
A nodal structure is dictated by symmetry in the case of $d_{x^2-y^2}$, $d_{xy}$ and $g$-wave, but in addition {\it there is a nodal structure close to the Brillouin zone diagonals, which is common for all solutions, regardless of the symmetry-imposed nodes}. We will return to this point below.

\begin{figure*}[t!]
\includegraphics[width=\linewidth]{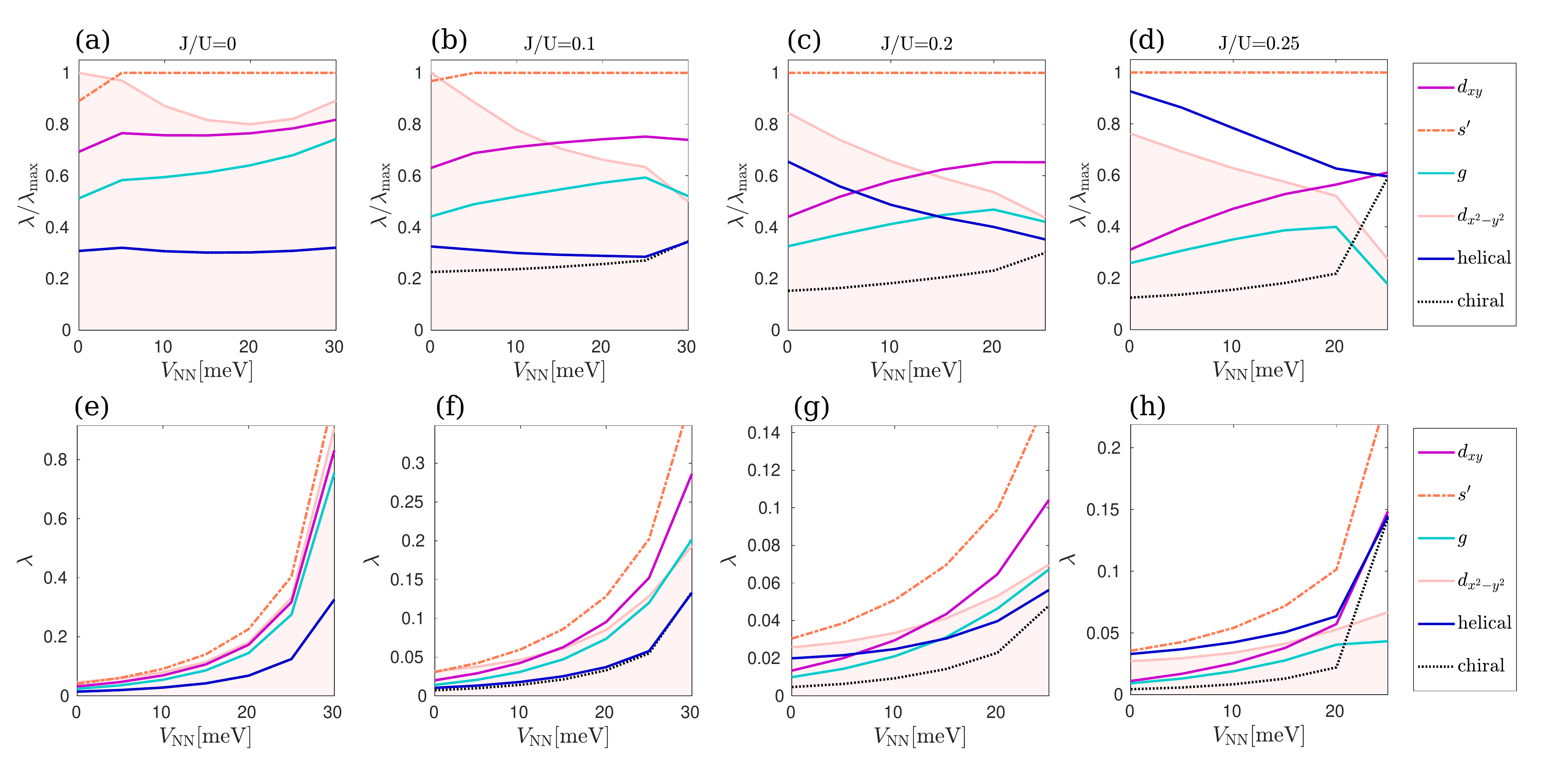}
\caption{Solutions to the linearized gap equation as a function of nearest-neighbor repulsion $V_\textrm{NN}$. Onsite Coulomb repulsion is $U=100$ meV and $\lsoc=45$ meV. Four different values of $J$ is chosen: (a) $J/U=0$ (note that in this case the helical and chiral solutions are degenerate and therefore the chiral solution is not visible), (b) $J/U=0.1$, and (c) $J/U=0.2$ and (d) $J/U=0.25$.
(a-d) For each value of $V_\textrm{NN}$, the eigenvalues of each irreducible representation is normalized to the maximal eigenvalue at that $V_\textrm{NN}$. The relative suppression of $d_{x^2-y^2}$ is clearly visible. (e-h) The evolution of the  eigenvalues of each symmetry channel on an absolute scale }
    \label{fig:DifferentHunds_SOC45}
\end{figure*}

In order to explore the robustness of the $s'$ and $d_{xy}$ near-degeneracy, we plot in Fig.~\ref{fig:DifferentHunds}(a-d) the {\it normalized} eigenvalues of Eq. (5) of the main text, where for each set of parameters we normalize all superconducting eigenvalues with respect to the leading eigenvalue. The eigenvalue of the leading instability within each irreducible representation, $s' ~(A_{1g})$, $d_{x^2-y^2} ~(B_{1g})$, $g=d_{x^2-y^2}d_{xy} ~(A_{2g})$, $d_{xy} ~(B_{2g})$ is depicted as a function of increasing nearest-neighbor Coulomb repulsion $V_\textrm{NN}$, keeping $\lsoc=35$ meV and $U=100$ meV constant while varying the value of the Hund's coupling $J/U$. 
The starting points of the calculation is chosen along the white line indicated in Fig.~\ref{fig:Phasedia}, which depicts the phase diagram of spin-fluctuation mediated pairing in \sruo~in the absence of longer-range Coulomb interactions~\cite{RomerPRL}.
In Fig.~\ref{fig:DifferentHunds} we observe the general trend of suppression of the $d_{x^2-y^2}$ solution relative to the other pairing channels as a function of $V_{\rm NN}$. This feature persists for all values of Hund's couplings. In the regime of small and moderate Hund's couplings $J/U <0.2$, the $s'$ and $d_{xy}$ solutions become the two leading solutions for nearest neighbor repulsion strengths $V_{\rm NN}>0.1U$.\\
For larger Hund's couplings we enter the regime where helical solutions are leading for  $V_\textrm{NN}=0$, see Fig.~\ref{fig:DifferentHunds} (d). In this regime, the helical solution persists as leading or subleading for all values of $V_{\rm NN}$ accessible before the proximity to a CDW instability renders the RPA formalism unreliable. 
The second leading solution in this case is $s'$ and therefore, for large Hund's couplings, the possibility of a $s'+ip$ solution appears, with $p$ being short-hand notation for one of the four helical solutions.\\
Fig.~\ref{fig:DifferentHunds} (e-h) show the eigenvalues of the different superconducting instabilities on an absolute scale. We point out the interesting fact that on an absolute scale, all superconducting channels undergo an increase as a function of increasing nearest-neighbor repulsion, as also seen in Fig. 1 of the main text.\\
To further test the robustness of the  $s'$ and $d_{xy}$ near-degeneracy induced by the presence of nearest-neighbor repulsion, we have investigated larger values of SOC as presented in Fig.~\ref{fig:DifferentHunds_SOC45}.
Also in this case, the $d_{x^2-y^2}$ solution is suppressed relative to the $s'$ and $d_{xy}$ channel. The latter two become leading for Hund's couplings $J/U=0.1-0.2$ and $V_{\rm NN} > 10 $ meV ($\simeq 0.1 U$). At larger Hund's coupling the helical solution becomes more competitive and for $J/U=0.25$ it remains subleading for nearest-neighbor repulsion up to $V_{\rm NN}=20$ meV. However, the $d_{xy}$ solution displays the strongest relative increase as a function of $V_{\rm NN}$ and therefore the $s'+id_{xy}$ might be favored even at large Hund's couplings when the nearest-neighbor repulsion becomes strong. \\
We note that spin-fluctuation mediated superconductivity does not find that $d_{x^2-y^2}$ and $g$-wave become the leading solutions at any value of $V_\mathrm{NN}$ in parameter space even though the two instabilities approach each other as a function of $V_{ \rm NN}$. \\

The superconducting instabilities obtained from spin-fluctuation mediated pairing deviate considerably from the simplest lowest-order harmonic gap functions of each irreducible representation. In order to quantify this, we have projected the linearized gap solutions $ \Delta (\kv)$ as seen from Fig.~\ref{fig:GapsThreeVnn} onto the lowest order harmonics $f(\kv)$ of the different irreducible representations:
\begin{equation}
    g_f=\int_{\kv\in FS}d\kv \Delta(\kv) f(\kv)/\int_{\kv\in FS}d\kv f(\kv)^2.
\end{equation}
The basis functions $f(\kv)$ are stated in Table~\ref{tab:Project} and the results of the projection procedure are shown for 
 the $s'$, $d_{xy}$ and $d_{x^2-y^2}$ solutions. 
\begin{table}[b!]
\centering
 \begin{tabular}{||c l | r|r| r||} 
 \hline
  & Basis function $f(\kv)$ & $V_{\rm NN}=0$  & $V_{\rm NN}=20$ meV  \\  
 \hline\hline
 $A_{1g}$ & 1 & -0.0002 & -0.0006  \\ 
 ($s'$)  & $\cos(k_x)+\cos(k_y)$    &  -0.0285     &  -0.0285      \\ 
   & $\cos(k_x)\cos(k_y)$      &  0.0575  &       0.0597  \\
   & $\cos(2k_x)+\cos(2k_y)$    & -0.0272 &       -0.0269 \\
   & $\cos(2k_x)\cos(2k_y)$     & 0.0012 &       -0.0128\\
   & $\cos(3k_x)+\cos(3k_y)$    & 0.0090 &      0.0127\\
   & $\cos(3k_x)\cos(3k_y)$     & 0.0067&      0.0129       \\
\hline
 $B_{2g}$ & $\sin(k_x)\sin(k_y)$        & -0.0115     &-0.0087 \\ 
 ( $d_{xy}$) & $\sin(2k_x)\sin(2k_y)$        & -0.0447&-0.0459\\ 
 & $\sin(3k_x)\sin(3k_y)$          &  0.0387 & 0.0427    \\ 
 & $\sin(4k_x)\sin(4k_y)$        &  -0.0178    & -0.0201   \\
\hline
 $B_{1g}$ & $\cos(k_x)-\cos(k_y)$        & 0.0240     & 0.0219\\ 
($d_{x^2-y^2}$) & $\cos(2k_x)-\cos(2k_y)$          & -0.0074  & -0.0061\\ 
 & $\cos(3k_x)-\cos(3k_y)$          & 0.0071 &    0.0101 \\ 
 & $\cos(4k_x)-\cos(4k_y)$        & -0.0009  &       -0.0011 \\
 & $\cos(5k_x)-\cos(5k_y)$        & -0.0101  &       -0.0109 \\
 & $\cos(6k_x)-\cos(6k_y)$        &  0.0132 &       0.0129 \\
 \hline
 \end{tabular}
 \caption{Coefficients of the projection to lowest order harmonics of the $s'$, $d_{xy}$ and $d_{x^2-y^2}$ as given by Eq.~(\ref{eq:lproj}). We have projected the linearized gap solutions in the case of $U=100$ meV, $J/U=0.1$ and $V_\textrm{NN}=0,20$ meV.}
 \label{tab:Project}
\end{table}
The bare nearest-neighbor repulsion will suppress pairing links for nearest neighbors, i.e. the channels $\cos(k_x)\pm\cos(k_y)$ of the $A_{1g}/B_{1g}$ channels.
From Table \ref{tab:Project}, it is seen that the main contribution to the $d_{x^2-y^2}$ superconducting instability is indeed strong links between electrons on neighboring sites, a feature which is unique for this pairing channel. Longer-range contributions to the pairing are, however, also significant even in the case of $V_{\rm NN}=0$. As $V_{\rm NN}$ increases, the nearest-neighbor links diminish and there is an increase in the third harmonics, as seen from Table 1.
Regarding the $s'$ solution, this solutions has strong pairing contributions at nearest and next-nearest neighbor sites, but even longer-range site separations also contribute in this pairing channel. In general, since the superconducting links are spread out more, the $s'$ solution will be relatively inert to the bare nearest-neighbor repulsion. Similar features apply for the $d_{xy}$ solution.
\begin{figure}[t!]
    \centering
        \includegraphics[width=0.9\linewidth]{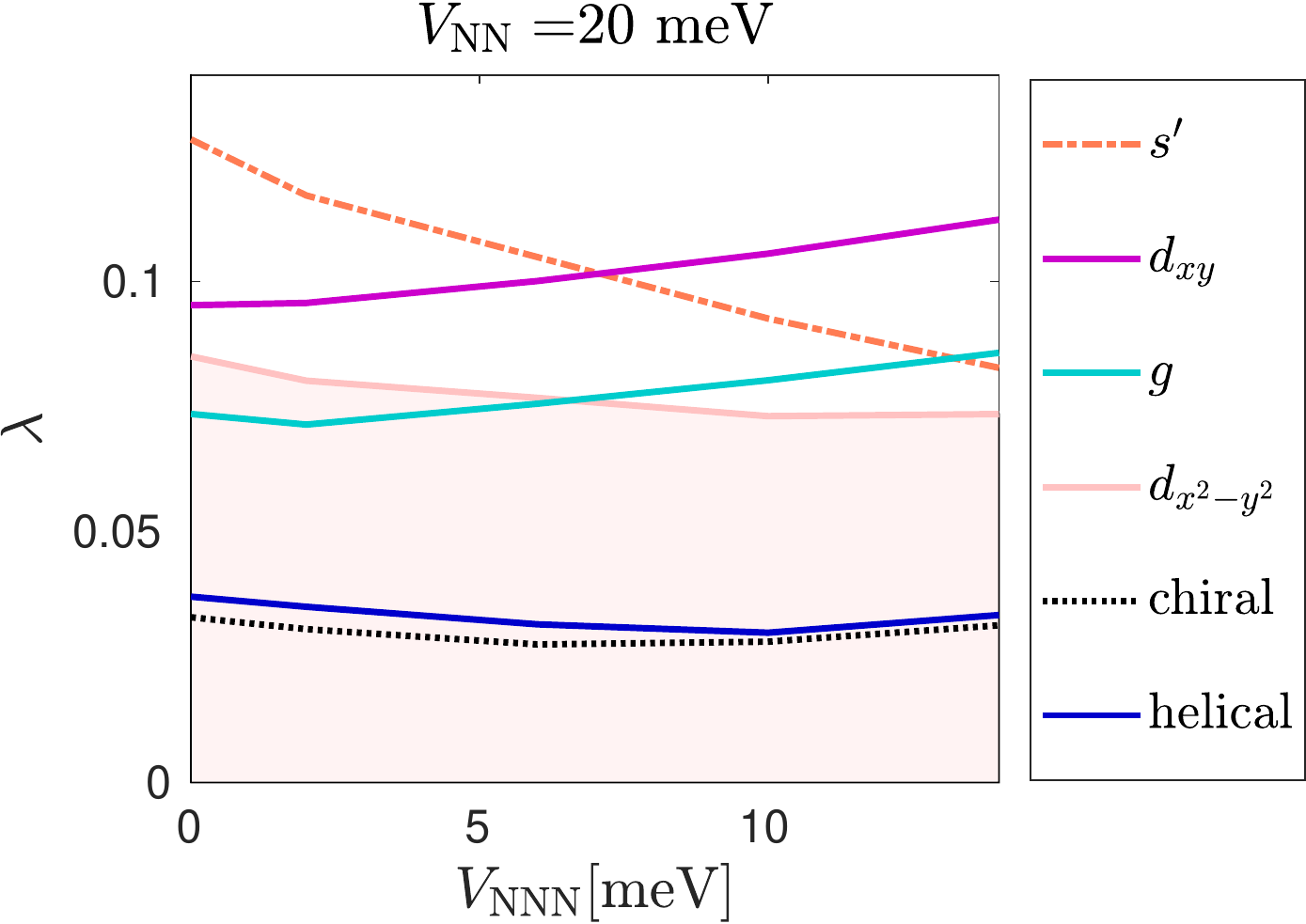}
\caption{Evolution of the superconducting instabilities as a function of {\it next}-nearest neighbor repulsion $V_{\rm NNN}$ in the case of nearest-neighbor repulsion $V_\textrm{NN}=20$ meV, $U=100$ meV, $J/U=0.1$ and $\lsoc=45$ meV.}
    \label{fig:NNN}
\end{figure}
From the projection coefficients presented in Table ~\ref{tab:Project}, we expect that a {\it next}-nearest neighbor repulsive interaction ($V_\textrm{NNN}$) will act destructively on the $s'$ channel (since the coefficient of the $\cos(k_x)\cos(k_y)$ basis function is largest for this solution), but the $d_{xy}$ channel should be relatively robust as the coefficient of $\sin(k_x)\sin(k_y)$ is subdominant to coefficients of the higher order harmonics.\\
Indeed, this is the case, as shown in Fig.~\ref{fig:NNN}, which reveals the effect of next-nearest-neighbor repulsion in the case of  $V_{\rm NN}=20$ meV  ($U=100$ meV and $J/U=0.1$). 
We observe that the near-degeneracy of $s'$ and $d_{xy}$ persists for increasing next-nearest-neighbor repulsion strength. In agreement with the expectations achieved from the projection procedure in Table 1,  the $d_{xy}$ solution is inert to next-nearest-neighbor repulsion (in fact there is even a small increase of the eigenvalue in this channel visible from Fig.~\ref{fig:NNN}), while the $s'$-solution becomes suppressed. However, the suppression of $s'$ is modest enough that $s'$ and $d_{xy}$ remain the two leading solutions for $V_{\rm NNN}<\frac{V_{\rm NN}}{\sqrt{2}}$ for the parameters used in Fig.~\ref{fig:NNN}.
\begin{figure*}
    \centering
        \includegraphics[width=\linewidth]{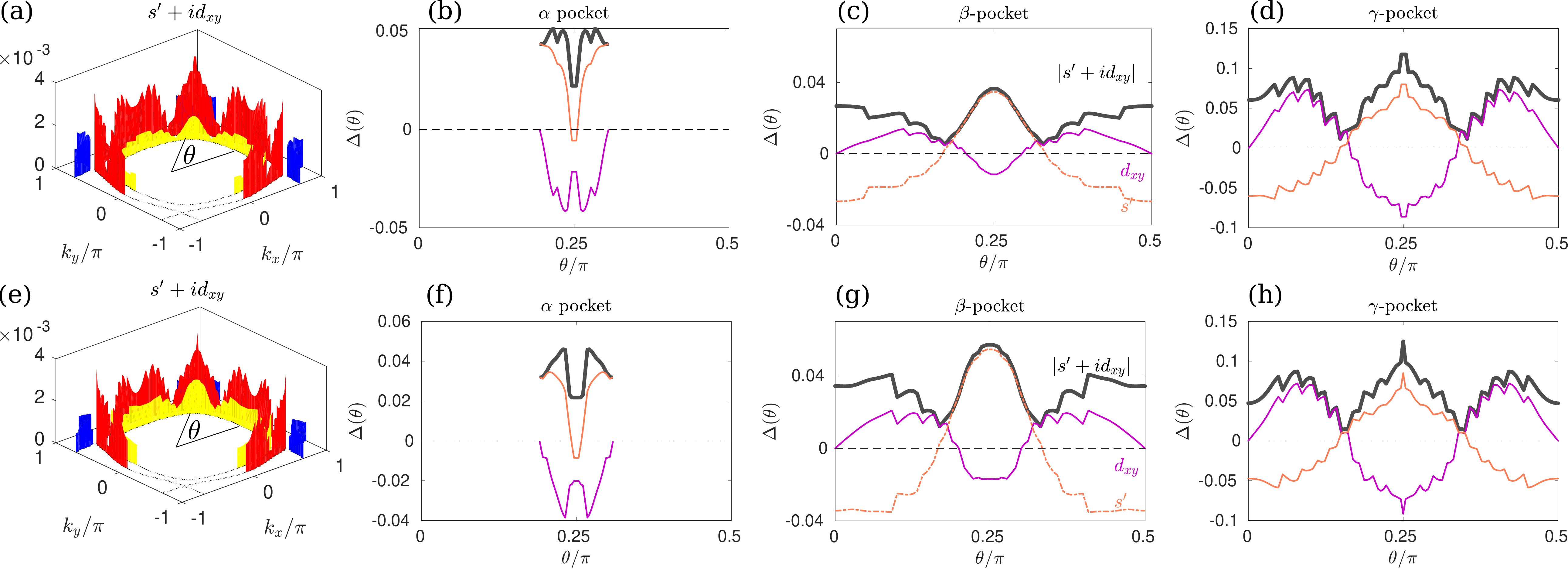}
\caption{(a,e) Spectral gap for the solution $s'+id_{xy}$ for nearest-neighbor repulsion $V_\textrm{NN}=20$ meV and $\lsoc=45$ and $\lsoc= 35$ meV, respectively.
(b-d) Superconducting gap at the $\alpha$-, $\beta$-, $\gamma$- pocket as a function angle $\theta$ defined in (a) for nearest-neighbor repulsion $V_\textrm{NN}=20$ meV, $U=100$ meV, $J/U=0.1$ and $\lsoc=45$ meV $=0.5t_1$. (f-h) depict the same as in (b-d) but with a decreased value of SOC to  $\lsoc=35$ meV ($=0.4t_1$).}
    \label{fig:AngleDependence}
\end{figure*}

\section{Nodal structure}
In Figs.~\ref{fig:DifferentHunds} and ~\ref{fig:DifferentHunds_SOC45}, the evolution of gap solutions belonging to different irreducible representations was depicted as a function of $V_\textrm{NN}$ for different values of Hund's couplings and SOC. Now we turn to a more detailed inspection of the gap structure in each case, and in particular we focus on the (near)-nodal structure of the combined solution $s'+i d_{xy}$.\\
Each gap solution will, with the exception of the $s'$ solution, display nodes, which are imposed by symmetry.
In addition, we find that another nodal structure appears to be common for all solutions irrespective of the symmetry. These nodes we refer to as nesting-enforced nodes since, as we shall see, for spin-fluctuation mediated pairing in \sruo~there is another source of nodal structure which is robust, and due to the strong nesting features and orbital character of the states at the Fermi surface. As a consequence, {\it these additional orbital-induced nodes occur at similar $\kv$ positions for solutions belonging to different irreducible representations}. This means that a near-nodal structure is present for superpositions of superconducting solutions belonging to different symmetry classes, unlike what is generally true in the case of symmetry-imposed nodes.
In order to visualize this property, we observe from in Fig.~\ref{fig:GapsThreeVnn} that the three leading solutions all have nodes close to the zone diagonal irrespective of whether the symmetry is $d_{x^2-y^2},s'$ or $d_{xy}$. The $g$-wave solution appears as the fourth leading solution for $V_{\rm NN}=20$ meV (not shown) and also have similar nodal structure in addition to the symmetry-enforced eight nodes.
 While superimposing two even-parity solutions  will in general lift the symmetry-imposed nodal structure (with the exception of $d_{x^2-y^2}+ig$), the nesting-enforced nodal structure remains.
 \begin{figure*}
    \centering
        \includegraphics[width=0.85\linewidth]{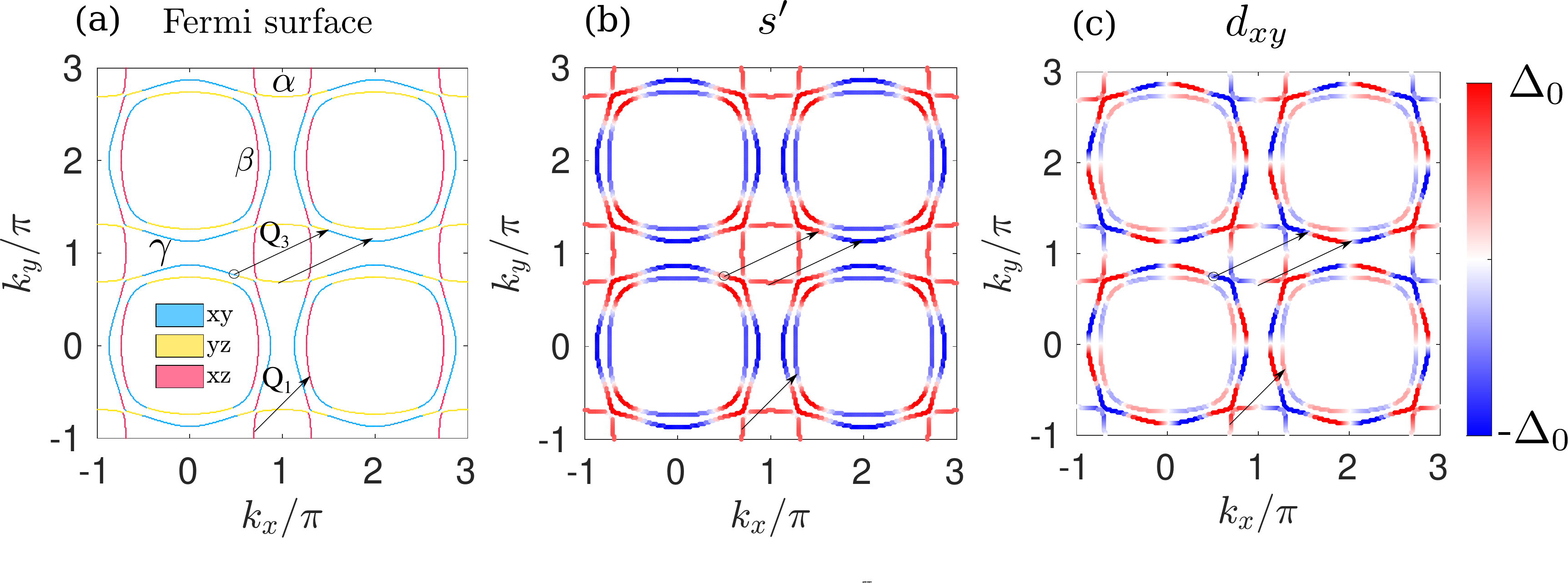}
        \caption{(a) The Fermi surface showing the dominant orbital character and the nesting vectors ${\bf Q}_1$ and ${\bf Q}_3$ . Note that while ${\bf Q}_1$ is mainly an intra-orbital nesting vector due of $xz-xz$ or $yz-yz$ orbital character, the ${\bf Q}_3$ nesting vector is of {\it inter}-orbital character ($xz-xy$ or $yz-xy$). The black circle marks the position where the dominating orbital content at the Fermi surface changes between $yz$ (yellow) and $xy$ (blue). (b,c)    Leading $s'$ and $d_{xy}$ gap solutions for $V_\mathrm{NN}=20$ meV, $U=100$ meV, $J/U=0.1$, and $\lsoc=35$ meV. The sign change in both gap functions found at the $\gamma$-pocket can be understood from inspection of the inter-orbital nesting vector ${\bf Q}_3$, as explained in the text.}
        \label{fig:GapAndNesting}
\end{figure*}
Specifically, we now discuss in detail the structure of the solution $s'+i d_{xy}$. This gap structure does not have any symmetry-imposed nodes. However, as outlined above, the multi-orbital structure of the pairing interaction favors a nodal structure in addition to symmetry-imposed nodes with additional nodes that are positioned in nearly the same place for $s'$ and $d_{xy}$. Therefore, this will lead to a near-nodal structure, even for complex superpositions like e.g. $s'+i d_{xy}$.  To zoom in on this property, we show in Fig.~\ref{fig:AngleDependence} the angle dependence of the individual gaps $s'$ and $d_{xy}$ as well as the gap magnitude of the complex superposition $s'+id_{xy}$. Within the resolution of the numerical calculation, we see near-nodes or deep minima in the combined gap solution, shown by the black line in Fig.
~\ref{fig:AngleDependence}(b,c,d) for $V_\textrm{NN}=20$ meV and $\lsoc=45$ meV and
Fig.
~\ref{fig:AngleDependence}(f,g,h) for $V_\textrm{NN}=20$ meV and $\lsoc=35$ meV. Especially, the $\gamma$-pocket shows a clear near-nodal structure.\\
We can understand the near-nodal structure at the $\gamma$-pocket as a consequence of $\Qv_3$-nesting, as we will now explain. In Fig.~\ref{fig:GapAndNesting} (a), we show the Fermi surface with dominating orbital character in the extended zone scheme, as well as the nesting features $\Qv_3$ and $\Qv_1$. Figure~\ref{fig:GapAndNesting} (b,c) show the two gaps $s'$ and $d_{xy}$. It is seen that the largest gap is found at the $\gamma$-pocket, even though all bands have a finite superconducting gap.
The sign change in both gap structures found at the $\gamma$-pocket can be understood from inspection of the inter-orbital nesting vector ${\bf Q}_3$.
Due to this nesting vector, the spin-fluctuation mediated pairing interaction will favor a sign-change between states residing at $\kv$ and $\kv+\Qv_3$. The $s'$ and $d_{xy}$ gaps are even parity and it is  therefore required to have nodes along the Fermi surface in order to accommodate this sign change while maintaining an even parity gap. The most favorable position of the nodes is in places where the $\Qv_3$ vector is inactive, i.e. in regions of like orbital character ($xy-xy$). This position is indicated in Fig.
~\ref{fig:GapAndNesting} by the left-most $\Qv_3$ arrow of each subfigure. Note that the $\Qv_3$ vector originating at the position of orbital content change points exactly at the node on the $\gamma$-pocket.
In addition to the importance of the $\Qv_3$ vector in the spin-fluctuation mediated pairing discussed in the previous section, we observe that the vector $\Qv_1$, which is of intra-orbital character, is sub-dominant regarding pairing. In the case of $s'$-wave superconductivity, the gap changes sign under $\Qv_1$, see Fig.~\ref{fig:GapAndNesting}(b), but in the case of the $d_{xy}$ gap, there is no sign-change between the gap at $\kv$ and $\kv+\Qv_1$, as indicated by the black arrow in the bottom of Fig.
~\ref{fig:GapAndNesting}(c). The gap magnitude on the $\beta$-pocket is small in order to avoid this unfavorable sign match.
\section{Knight shift calculation}
\begin{figure}[t!]
    \centering
        \includegraphics[width=\linewidth]{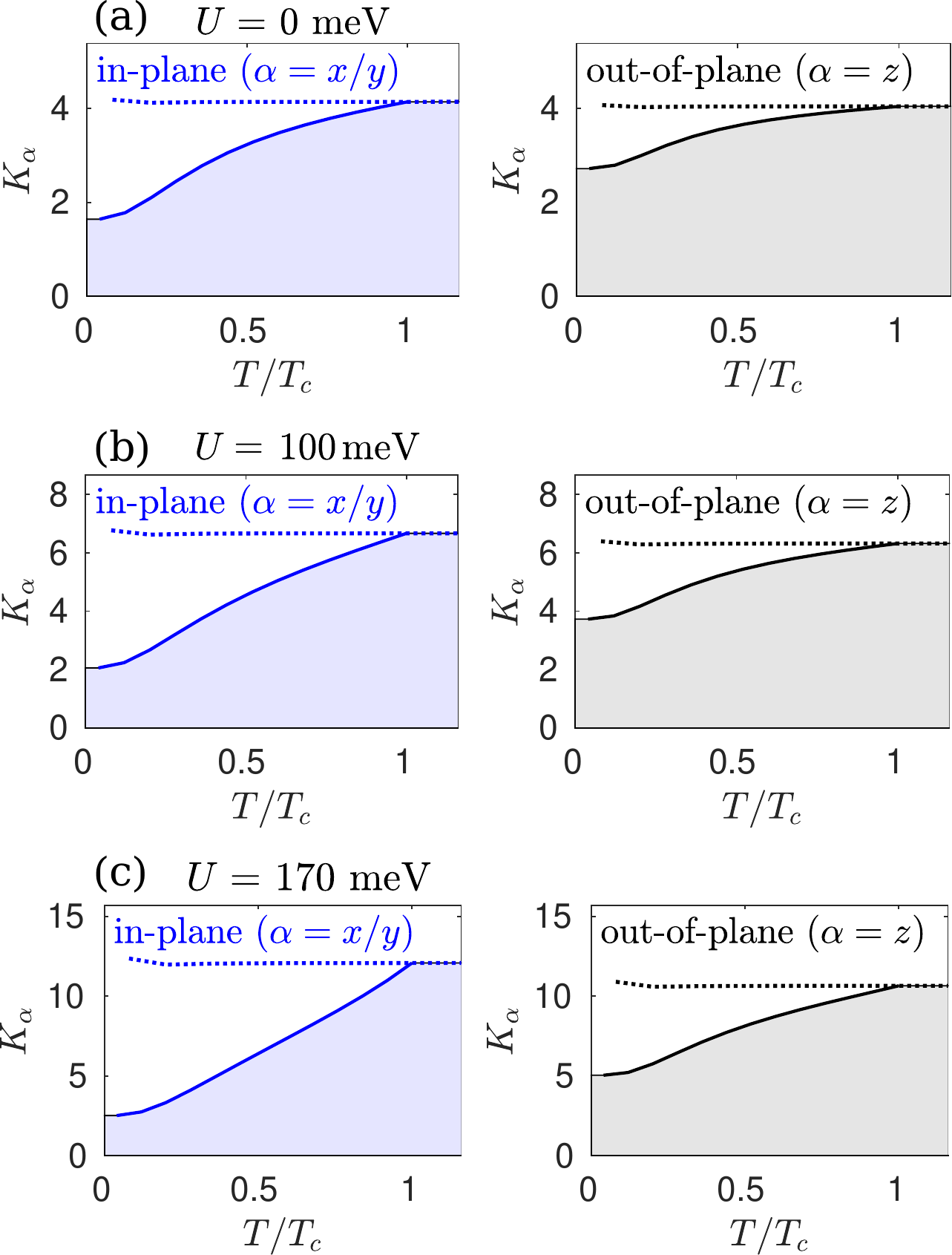}
\caption{Knight shift as calculated by Eq.~(\ref{eq:Ka}) for the solution $s'+id_{xy}$. The superconducting solutions are obtained for $U=100$ meV, $J/U=0.1$ and $V_{\rm NN}=20$ meV for $\lsoc=35$ meV. The RPA susceptibility is calculated as stated in Eq.~(\ref{eq:chiRPA}) for three different values of $U=0,100,170$ meV for (a,b,c), keeping $J/U=0.1$ and $V_{\rm NN}/U=0.2$ in all cases. In-plane magnetic fields are shown in the left column ($\alpha=x,y$) and out-of-plane magnetic fields are shown in the right column where full (dashed) lines depict the superconducting (normal) state response. The maximum gap value at $T=0$ has been set to $\Delta=1.5$ meV and $T_{c}=0.5$ meV.}
    \label{fig:KS_SOC35}
\end{figure}
Finally, we discuss how to obtain a Knight shift response within our framework. As presented in the main text, we calculate the Knight shift 
in the linear response regime, i.e. we do not include an explicit magnetic field in the calculation and the results therefore corresponds to the very weak-field limit. For an external magnetic field along $ \alpha \in\{x,y,z\}$, the Knight shift response is obtained from the static spin-resolved susceptibility by
\begin{equation}
K_\alpha =\frac{1}{4} \sum_{\mu,\nu} \spin_{s_1,s_2}^\alpha \spin_{s_3,s_4}^\alpha [\chi_{\rm sc}]^{\mu, s_1; \mu,s_2}_{\nu, s_3;\nu s_4},
\label{eq:Ka}
\end{equation}
where $\spin^\alpha$ denotes the Pauli matrices, and 
\begin{eqnarray}
\Big[\chi_{\rm sc}\Big]^{\io_1,\io_2}_{\io_3,\io_4}&=&  \Big[\frac{1}{1-\chi_{\rm sc,0}W}\chi_{\rm sc,0} \Big]^{\io_1,\io_2}_{\io_3,\io_4}(\qv,i\omega_n,\delta,\delta'),
\label{eq:chiRPA}
\end{eqnarray}
is the RPA spin susceptibility in the superconducting state evaluated at $\qv=(0,0)$, $i\omega_n=0$ and $\delta=\delta'=0$. Summation over spins, orbitals and lattice vectors $\delta$ is implicit in Eq.~(\ref{eq:chiRPA}).\\
The Knight shift $K_\alpha$ is shown for in-plane $(\alpha=x,y)$ and out-of-plane $(\alpha=z)$ magnetic field direction in Fig.~\ref{fig:KS_SOC35}  for the $s'+id_{xy}$ solution obtained for $V_{\rm NN}=20$ meV ($U=100$ meV and $J/U=0.1$). We find that the magnitude of the suppression of $K^\alpha$ deep in the superconducting state compared to the normal state value depends somewhat on the interaction parameters chosen in the RPA calculation, Eq.~(\ref{eq:chiRPA}). We compare the case of $U=0,100,170$ meV with the ratios $J/U=0.1$ and $V_{\rm NN}/U=0.2$ kept constant. In the case of $U=170$ meV we get a Knight shift at the lowest temperature which is roughly $20 \%$ of the normal state value for in-plane fields, while out-of-plane fields give a smaller suppression of approximately $50 \%$. Further increase of $U$ will give rise to an even more pronounced decrease in both channels as can also be inferred from the tendencies in Fig.~\ref{fig:KS_SOC35}. Note that it is the presence of SOC which renders the in-plane and out-of-plane Knight shift suppression different in the case of even-parity solutions.
For in-plane fields a large suppression of the Knight shift was recently reported in Ref.~\onlinecite{Chronister}. The results of the Knight shift suppression due to $s'+id_{xy}$ superconductivity presented in Fig.~\ref{fig:KS_SOC35} are in rough agreement with the experimental observation, but we point out that the quantitative suppression result depends on the interaction strength as discussed above.
\begin{figure}[t!]
    \centering
        \includegraphics[width=\linewidth]{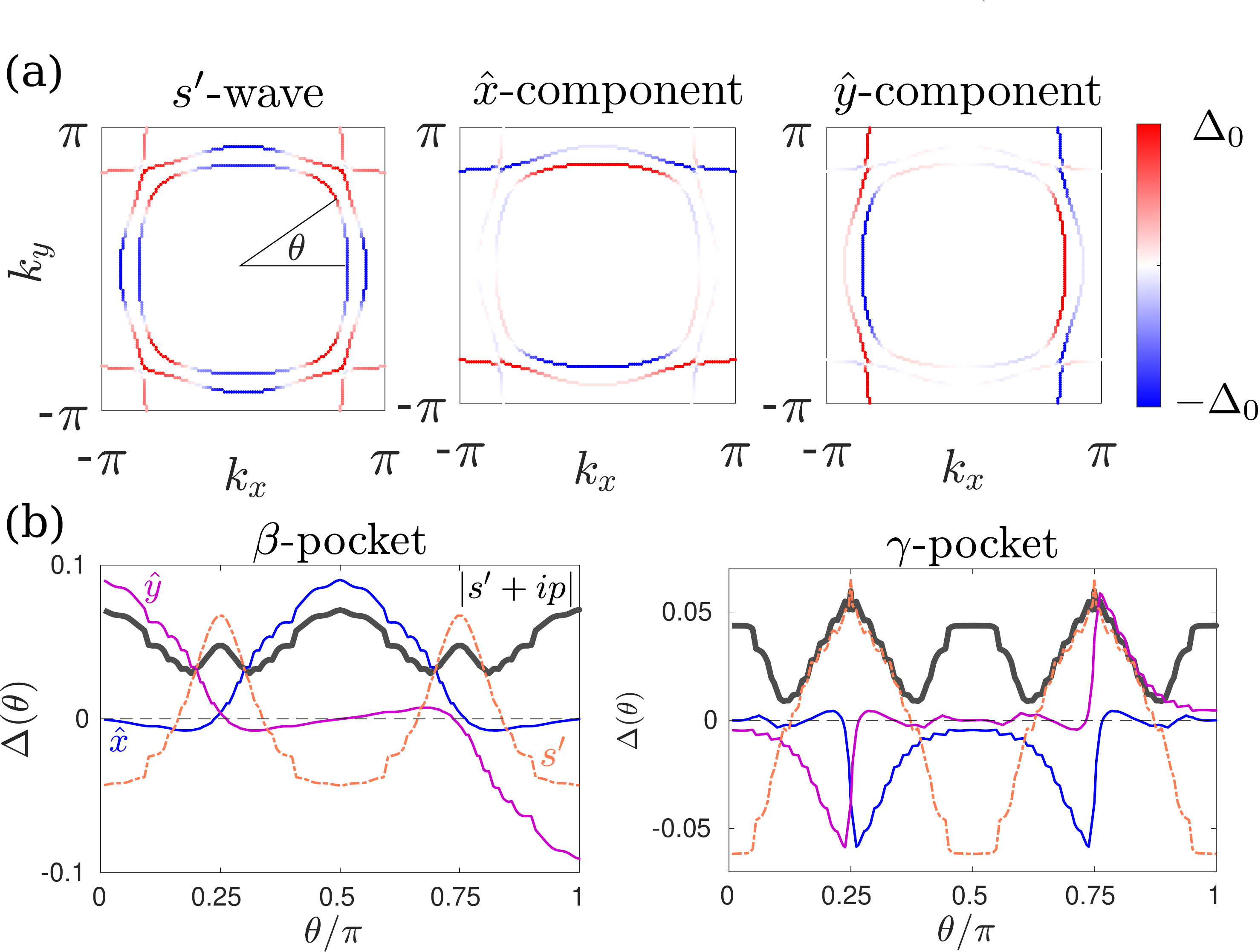}
\caption{(a) The two leading gap solutions ($s'$-wave and helical structure with $\hat x$ and $\hat y$ components shown separately ) in the case of large Hund's coupling $J/U=0.25$ for onsite Coulomb interaction $U=100$ meV, nearest-neighbor repulsion $V_\textrm{NN}=20$ meV and spin-orbit coupling $\lsoc=35$ meV. The $s'$ solution is the leading instability, while the helical solution is second leading. All four helical solutions are degenerate in this model. (b) The gap at the Fermi surface as a function of the angle $\theta$ as defined in (a). The gap is shown for the separate components $s'$ (dashed orange), $\hat x$ (blue), $\hat y$ (magenta) and the combined solution $s'+ip$ (black) , where $p=\Delta_x \hat x + \Delta_y \hat y $ denotes one of the helical solutions.}
\label{fig:SIP}
\end{figure}
\section{Large Hund's couplings and helical solutions}
\begin{figure}[b!]
    \centering
        \includegraphics[width=\linewidth]{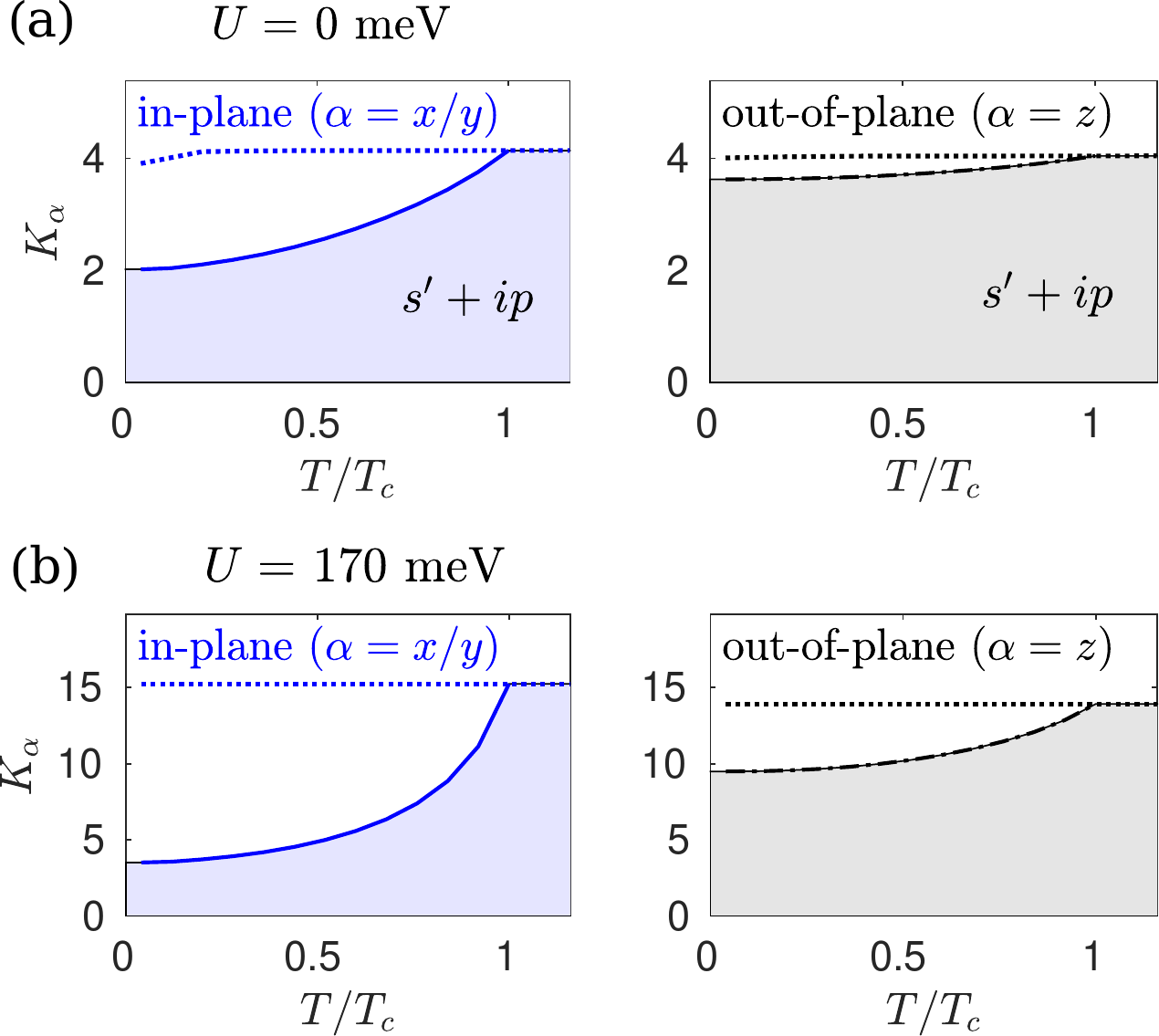}
\caption{Knight shift as calculated by Eq.~(\ref{eq:Ka}) for the solution $s'+ip$ where $p=\Delta_x \hat x + \Delta_y \hat y $ denotes one of the helical solutions for in-plane (left) and out-of-plane (right) fields, respectively. The superconducting solutions are obtained for $U=100$ meV, $J/U=0.25$ and $V_{\rm NN}=20$ meV with $\lsoc=35$ meV. The depicted data corresponds to (a) the bare susceptibility, i.e. $U=J=V_{\rm NN}=0$ in the RPA expression of Eq.~(\ref{eq:chiRPA}) and
(b) the RPA susceptibility taking $U=170$ meV $J/U=0.25$  and $V_{\rm NN}=0$ in the RPA expression of Eq.~(\ref{eq:chiRPA}).}
\label{fig:KS_sip}
\end{figure}
To investigate the effect of the Hund's coupling $J$, we addressed in Figs.~\ref{fig:DifferentHunds} and~\ref{fig:DifferentHunds_SOC45}  four different values of the Hund's coupling $J/U=0,0.1,0.2,0.25$. In the regime of large Hund's couplings, the helical solutions become competitive to the subleading $d_{xy}$ (focusing on $V_{\rm NN}=15-25$ meV), while $s'$ remains leading.
The possibility of a TRSB superconducting instability of the form $s'+ip$, where $p$ denotes one of the helical solutions, thus becomes relevant. The three constituent parts of such a solution is shown in Fig.~\ref{fig:SIP}(a).
While previous calculations in the limit of $V_\textrm{NN}=0$ presented in Ref.~\cite{RomerPRL} found that the helical solution features a fully gapped superconducting state this is not the case for the current gap structures (for finite $V_\textrm{NN}=20$ meV) since there is a tendency of a washed out gap at the $\gamma$-pocket close to $(\pi,0)/(0,\pi)$ as seen from the $\bf \hat x$ and $\bf \hat y$ gap components in Fig.~\ref{fig:SIP}.
This points to a preservation of a near-nodal gap even in the case of the gap of type $s'+ip$, where $p$ is used as short-hand notation for a helical gap. We plot in Fig.~\ref{fig:SIP}(b) the gap at the Fermi surface pockets $\beta$ and $\gamma$ for the $s'+ip$ gap as well as its constituents, and observe that there is indeed a near-nodal structure on the $\gamma$-pocket close to the zone diagonals.\\
This observation makes it worthwhile to consider the Knight shift suppression in the superconducting state in the case of the exotic mixed-parity solution $s'+ip$. We have performed this calculation with the result shown in Fig.~\ref{fig:KS_sip} for in-plane and out-of-plane field directions. The curves in Fig.~\ref{fig:KS_sip}(a) show the Knight shift results for the bare susceptibility. For in-plane fields we obtain a very similar result compared to the bare susceptibility result displayed in Fig.~\ref{fig:KS_SOC35}(a). Thus, depending on the choice of interaction parameters in the RPA calculation of Eq.~(\ref{eq:chiRPA}), the $s'+ip$ solution could in fact be in agreement with the experimental result of Ref.~\onlinecite{Chronister} as shown in Fig.~\ref{fig:KS_SOC35}(b) for $U=170$ meV and $J/U=0.25$ ($V_{\rm NN}=0$).\\
Out-of-plane fields provide a means of distinguishing between the $s'+id_{xy}$ and $s'+ip$ scenario, since there as an almost insignificant suppression of the Knight shift for out-of-plane field values in the case of $s'+ip$ owing to the helical component while the  $s'+id_{xy}$ gap gives rise to a more pronounced suppression for out-of-plane fields.
%

\end{document}